# Finding substructures in protostellar disks in Ophiuchus

Arnaud Michel ,[1] Sarah I. Sadavoy ,[1] Patrick D. Sheehan ,[2] Leslie W. Looney ,[3] Erin G. Cox ,[4]
John J. Tobin ,[5] Nienke van der Marel ,[6] and Dominique M. Segura-Cox [7]

[1] *Department of Physics, Engineering Physics and Astronomy, Queen's University, Kingston, ON, K7L 3N6, Canada*
[2] *National Radio Astronomy Observatory, 520 Edgemont Rd., Charlottesville, VA, 22903, USA*
[3] *Department of Astronomy, University of Illinois, Urbana, IL 61801, USA*
[4] *Center for Interdisciplinary Exploration and Research in Astronomy, Northwestern University, 1800 Sherman Rd., Evanston, IL 60202, USA*
[5] *National Radio Astronomy Observatory, Charlottesville, VA, USA*
[6] *Leiden Observatory, Leiden University, P.O. Box 9513, 2333 CA, Leiden, The Netherlands*
[7] *Department of Astronomy, The University of Texas at Austin, 2515 Speedway, Austin, TX 78712, USA*

## ABSTRACT

High-resolution, millimeter observations of disks at the protoplanetary stage reveal substructures such as gaps, rings, arcs, spirals, and cavities. While many protoplanetary disks host such substructures, only a few at the younger protostellar stage have shown similar features. We present a detailed search for early disk substructures in ALMA 1.3 and 0.87 mm observations of ten protostellar disks in the Ophiuchus star-forming region. Of this sample, four disks have identified substructure, two appear to be smooth disks, and four are considered ambiguous. The structured disks have wide Gaussian-like rings ($\sigma_R/R_{\rm disk} \sim 0.26$) with low contrasts ($C < 0.2$) above a smooth disk profile, in comparison to protoplanetary disks where rings tend to be narrow and have a wide variety of contrasts ($\sigma_R/R_{\rm disk} \sim 0.08$ and $C$ ranges from $0-1$). The four protostellar disks with the identified substructures are among the brightest sources in the Ophiuchus sample, in agreement with trends observed for protoplanetary disks. These observations indicate that substructures in protostellar disks may be common in brighter disks. The presence of substructures at the earliest stages suggests an early start for dust grain growth and, subsequently, planet formation. The evolution of these protostellar substructures is hypothesized in two potential pathways: (1) the rings are the sites of early planet formation, and the later observed protoplanetary disk ring-gap pairs are secondary features, or (2) the rings evolve over the disk lifetime to become those observed at the protoplanetary disk stage.

## 1. INTRODUCTION

The advent of high-resolution millimeter observations with the Atacama Large Millimeter/submillimeter Array (ALMA) has demonstrated that protoplanetary disks can host significant substructures (e.g., ALMA Partnership et al. 2015; Andrews 2020). Substructures range from rings, gaps, arcs, to spirals and have been observed in dust and CO isotopologues (e.g., ALMA Partnership et al. 2015; Andrews et al. 2016; Isella et al. 2016; Cox et al. 2017; Andrews et al. 2018; Huang et al. 2018; Long et al. 2018; Francis & van der Marel 2020; van der Marel et al. 2021; Teague et al. 2021; Pinte et al. 2022). The findings of different substructure features have motivated many diverse theoretical substructure-forming mechanisms (see review by Andrews 2020), including envelope infall (e.g., Bae et al. 2015; Kuznetsova et al. 2022), zonal flows (e.g., Johansen et al. 2009; Uribe et al. 2011; Bai & Stone 2014), dead zones (e.g., Regály et al. 2012; Dzyurkevich et al. 2013), snow lines (e.g., Stevenson & Lunine 1988; Stammler et al. 2017; Owen 2020), photoevaporative clearing (e.g., Clarke et al. 2001; Alexander et al. 2014; Ercolano & Pascucci 2017),

Corresponding author: Arnaud Michel
arnaud.michel@queensu.ca

gravitational instabilities (e.g., Toomre 1964; Kratter & Lodato 2016), gravitational perturbations from a companion (e.g., Cuello et al. 2019; Offner et al. 2022), and planets (e.g., Lin & Papaloizou 1979; Goldreich & Tremaine 1980; Dong et al. 2015; Paardekooper et al. 2022).

Disk substructure has been primarily studied in large and nearby ($\lesssim$150 pc) protoplanetary disks, which are generally envelope-less, optically thin (ring-like substructures may be optically thick), and there is a large contrast between the gaps and rings, making it easier to directly detect substructures at high resolution (ALMA Partnership et al. 2015; Andrews et al. 2018; Huang et al. 2018). From an observational perspective of protoplanetary disk demographics, van der Marel & Mulders (2021) find that the stellar mass, disk substructures, and giant planet occurrence are closely linked. This study suggests that the correlation between stellar mass and significant disk substructures leads to higher occurrence rates of giant planet formation. While the properties of the host disk likely dictate the types of planets that will ultimately form, the formation of a giant planet can greatly influence the disk's longevity and the formation of any additional planets (e.g., Mulders et al. 2021a; Michel et al. 2021). However, it is also likely that the type of planet that is first formed is linked to the original disk properties.

If large planets induce substructures in protoplanetary disks, the planet formation process, interconnected with substructure formation, must start earlier during the protostellar phase (Tychoniec et al. 2020; Cridland et al. 2021; Miotello et al. 2022). Detecting substructures in embedded protostellar disks is more challenging due to confusion from the surrounding envelope and higher disk optical depths. Nevertheless, recent ALMA observations have discovered annular substructures in some protostellar disks, e.g., IRS 63 (Segura-Cox et al. 2020) and a series of disks in Orion (Sheehan et al. 2020). IRS 63 hosts two annular rings on top of the disk profile that are more like annular flux enhancements than distinct rings (Segura-Cox et al. 2020). The structured protostellar disks in Orion are significantly different, with cavities and defined rings more akin to transition disks (e.g., Francis & van der Marel 2020). Furthermore, warps, spirals, flaring, and asymmetric features have also been found in some protostellar disks (e.g., Tobin et al. 2016; de Valon et al. 2020; Ohashi et al. 2022; Sheehan et al. 2022b; Michel et al. 2022; Lin et al. 2023; Yamato et al. 2023).

This paper presents four structured protostellar disks from a sample of ten Ophiuchus protostellar sources. Three structured disks are new detections, and the fourth is IRS 63. In Section 2, we present the ALMA 1.3 and 0.87 mm observations of the protostellar disk sample and describe the methods used to model the disks and identify substructures from the *uv* visibilities. In Section 3, we present the results and categorize the disks as structured, non-structured, and ambiguous. In Section 4, we examine the structured disk properties in the context of theoretical predictions for disk evolution and planet formation, and we compare these disks to other known structured protostellar and protoplanetary disks. In Section 5, we present our conclusions and different avenues for future work.

## 2. METHODS

### 2.1. *Sample selection*

Ophiuchus is a nearby ~139 pc (Ortiz-León et al. 2018; Esplin & Luhman 2020), young, and active star-forming region (Wilking et al. 2008; Esplin & Luhman 2020; Gupta & Chen 2022). The proximity of the region has enabled high-resolution (Cieza et al. 2021), sensitive (Sadavoy et al. 2019), and complete disk sample (Esplin & Luhman 2020) observations at different millimeter wavelengths. There are 25 Class 0/I individual protostellar sources with very high sensitivity ALMA observations at 1.3 mm or 0.87 mm (Harris et al. 2018; Sadavoy et al. 2019). In Table 1, we outline the ALMA data sets used and their details. The data sets are self-calibrated as described in Harris et al. (2018), Sadavoy et al. (2019), and Encalada et al. (2021). We averaged the datasets to a single spectral channel and in 30s time bins. The reported beam sizes and rms values are obtained by imaging the sources using `tclean` with Briggs robust of 0.5. Most sources were observed at the phase center of the primary beam, where flux calibration intrinsically has a 5-10% uncertainty. Additionally, a few sources (e.g., GSS 30 IRS 3) were significantly offset and thus are primary beam corrected during the imaging process.

We selected the most massive single or wide (>350 au) binary sources from the high-sensitivity disk samples to search for early evidence of substructure. This corresponds to disks with a total mass > 1.8 $M_{\rm Jup}$[1] and a peak signal-to-noise > 300 at 1.3 mm. For simplicity of analysis, we excluded the close (<150 au) binary systems, IRS 43, IRS 67, and VLA 1623 A/B. The selection of isolated sources with deep observations allows us to search for subtle variations in the protostellar disk sub-

[1] Toal masses are obtained assuming a dust opacity of 3.5 cm$^2$g$^{-1}$ at 345 GHz, a temperature of 20 K, and a dust-to-gas ratio of 1:100.

**Table 1.** Programs observing the Ophiuchus protostellar disks

| ALMA Band | $\lambda$ (mm) | Program | PI | Beam ($''$) | rms ($\mu$Jy) | rms ($M_\oplus$)[c] | Frequency (GHz) | Baselines[a] (m) | MRS[b] ($''$) |
|---|---|---|---|---|---|---|---|---|---|
| 6 | 1.3 | 2015.1.01112.S | S. Sadavoy | 0.27 × 0.21 | 30 | 1.5 | 233 | 17-2647 / 15-1124 | 1.4 / 2.6 |
| 7 | 0.87 | 2015.1.00741.S | L. Looney | 0.15 × 0.11 | 220 | 4 | 345 | 15-612 / 15-460 | 1.4 / 7.2 |
| 7 | 0.87 | 2015.1.00084.S[d] | L. Looney | 0.16 × 0.15 | 110 | 2 | 345 | 14-1450 | 2.3 |

*Notes:*

a Two sets of baseline ranges are reported due to the respective observation series.

b Maximum Recoverable Scale (MRS). Two values are reported corresponding to the respective observation blocks.5

c The mass sensitivity evaluated is based on an assumption of optically thin dust, using Equation (8) from Sadavoy et al. (2019). We use typical values such that we have a distance of $\sim$ 139 pc, a dust temperature of 20 K, and $\kappa_d = 2.4$ cm$^2$g$^{-1}$ at 233 GHz and $\kappa_d = 3.5$ cm$^2$g$^{-1}$ at 345 GHz (Andrews et al. 2009).

d Only one source in our sample, VLA 1623W, was observed as part of this program.

**Table 2.** Ophiuchus protostellar disk sample

| Source | Class | System | $\text{Peak}_{1.3\text{mm}}$ (mJy) | $\sigma_{1.3\text{mm}}$ ($\mu$Jy) | $\text{Peak}_{0.87\text{mm}}$ (mJy) | $\sigma_{0.87\text{mm}}$ ($\mu$Jy) | $T_{\text{bol}}$[a] (K) | $L_{\text{bol}}$[b] ($L_\odot$) |
|---|---|---|---|---|---|---|---|---|
| Elias 29 | I | Single source | 15.8 | 36 | 32.9 | 282 | 350 | 18 |
| GSS 30 IRS 1 | I | Wide binary | 12.5 | 36 | 25.1 | 292 | 300 | 8.7 |
| GSS 30 IRS 3 | I | Wide binary | 39.5 | 92 | 52.5 | 1998 | 300 | 0.13 |
| IRS 37-A | I | Wide multiple | 9.7 | 32 | 17.4 | 236 | 130 | 2.6 |
| IRS 44 | I | Single source | 10.7 | 33 | 20.9 | 213 | 180 | 15 |
| IRS 63 | I | Single source | 68.7 | 71 | 103.9 | 226 | 530 | 1.4 |
| Oph-emb-1 | 0 | Single source | 10.4 | 30 | 23.8 | 222 | 49 | 0.3 |
| Oph-emb-6 | I | Single source | 17.8 | 35 | 31.1 | 222 | 170 | 0.2 |
| Oph-emb-9 | I | Single source | 222.4 | 34 | 40.6 | 226 | 280 | >0.1 |
| VLA 1623 West[c] | 0/I | Unknown[d] | 12.0 | 54 | 22.9 | 110 | 120 | >0.04 |

*Notes:*

a Bolometric temperature is provided as an independent indicator of youth, data from Evans et al. (2009).

b Bolometric luminosity is provided as a proxy for stellar mass (Dunham et al. 2014a; Fischer et al. 2017), data from Evans et al. (2009), except GSS 30 IRS 3, which is from Friesen et al. (2018).

c The VLA 1623 West field contains VLA 1623 Aa, Ab, and B. While these sources were removed from the visibilities (see Michel et al. 2022), some residual emission remains.

d The exact nature of VLA 1623 West relative to its nearby companions VLA 1623 Aa, Ab, B system is under debate. Harris et al. (2018) consider it to be an ejected source from the triple system, whereas the source could also be an unrelated object or non-coeval with A and B (Murillo et al. 2016). It is nevertheless found to be spatially and kinematically connected with A and B through accretion streamers (Mercimek et al. 2023).

structures (e.g., Gulick et al. 2021; Michel et al. 2022). Table 2 presents the protostellar disk sample.

To identify substructures in these disks, we model the *uv* visibilities (see Section 2.2). Before this, we corrected the *uv* visibility weights following Sheehan et al. (2020). The visibility weights should follow $\sigma_{vis} = \sqrt{1/\Sigma w_i}$, where $w_i$ is the weight of each visibility and $\sigma_{vis}$ is the root mean square of the naturally weighted image. In this work, the datasets require a 0.25 scaling factor for the visibility weights at both wavelengths so the noise matches the measured values from the imaging. This scaling factor is consistent with the scale factor found by Sheehan et al. (2020) and Michel et al. (2022).

### 2.2. *Disk models*

This work will refer to substructures in protostellar disks as gaps and rings over an underlying smooth disk model. A variety of analytic profiles for geometric disk models are available to fit *uv* visibilities (e.g., see Tazzari et al. 2021, for a brief list). We use analytic 1D brightness profiles to describe the protostellar source emission. For the fitting process, we assume that we have geometrically thin protostellar disks with no verti-cal height. We also assume that the emission can be well fit with an axisymmetric brightness profile (e.g., Pinilla et al. 2021; Tazzari et al. 2021; Sheehan et al. 2022a). We focus on models using standard Gaussians, modified Gaussians with flat tops, and a Power Law Core with an exponential Tail (PLCT).

The Gaussian profile is a model frequently applied to fit a variety of astrophysical observations flux distributions, including millimeter disk observations (e.g., Ansdell et al. 2016, 2020; Tazzari et al. 2021). The profile used it,

$$I(R) = I_0 \exp\left(-0.5\left(\frac{R}{\sigma}\right)^{\phi}\right) \tag{1}$$

where $I(R)$ is the intensity as a function of radius $R$, $I_0$ is the peak intensity at the center, $\sigma$ is the standard deviation width, and $\phi$ is the exponent dictating the rate of the radial drop. An exponent of $\phi = 2$ is for a regular Gaussian, but we can also recover a modified *Flat-Topped Gaussian* (FTG) model when $\phi$ is a free parameter greater than 2. This sharper-edged model fits emission from objects with high inclinations and optical depths as these appear to have a constant flux along the major axis (e.g., Michel et al. 2022). We also employ a large-scale Gaussian to represent the surrounding envelope structure and fit the shortest $uv$-distances for some more embedded sources. For the large-scale Gaussian envelope component, we assume that it is spherical and fix the inclination and position angle to 0°. For the envelope profiles, we do not account for any primary beam attenuation; however, most envelopes are compact and toward the phase center, which will make this correction mostly negligible.

The PLCT profile is motivated by viscous disk accretion theory (Lynden-Bell & Pringle 1974) where the two-component power law and exponential tail can describe embedded protostellar disks (Andrews et al. 2009; Segura-Cox et al. 2020). We use this profile and interpret the components such that it roughly represents the inner protostellar disk, and the exponential trail addresses the edge of the outer disk and the inner envelope component. The PLCT profile is given by,

$$I(R) = I_0\left(\frac{R}{R_c}\right)^{-\gamma} \exp\left(-\left(\frac{R}{R_c}\right)^{2-\gamma}\right) \tag{2}$$

where $I(R)$ is the intensity as a function of radius $R$, $I_0$ is the peak intensity at the center, and $\sigma$ is the standard deviation width, $R_c$ is the characteristic protostellar disk radius and $\gamma$ is the surface density gradient. When the PLCT exponent $\gamma = 0$, we recover a standard Gaussian.

High-resolution observations of protoplanetary disks reveal a variety of substructures, including rings, gaps, asymmetries, and spirals (Andrews et al. 2018; van der Marel et al. 2021). Some of these features can be analytically described using simple Gaussians that are offset from the source center (e.g., Segura-Cox et al. 2020) as,

$$I(R) = \pm I_G \exp\left(-0.5\left(\frac{R - \mathrm{loc}_X}{\sigma}\right)^2\right) \tag{3}$$

where the $\pm$ variation allows us to simulate a ring (+) or a gap (−), $I_G$ is the peak intensity at the center of the annular feature, and $\mathrm{loc}_X$ is the offset from the disk center on which the Gaussian structure is centered.

In Table 3, we present the free parameters we use to define the analytic functions. For each source, we fit the model free parameters and the inclination $i$ of the disk along the line of sight, the position angle $P.A.$, and the source offsets from the field center ΔR.A. and ΔDec. The visibility profiles and imaged observations, models, and residuals, are then modelled based on this source center.

We have parameter spaces that range from 6-dimensional to 15-dimensional depending on the model and the number of substructures. We explore these parameter spaces with a Bayesian approach using an affine-invariant Markov chain Monte Carlo (MCMC) ensemble sampler, `emcee` v 2.2.1 (Foreman-Mackey et al. 2013), within the `GALARIO` python package (Tazzari et al. 2018). We find the optimal fit parameters using 60 walkers over 5000 steps distributed using `MPIPool` for a $R = 5''$ grid with a $10^{-5}$ arcsecond cell size. This setup provides quick convergence without sacrificing the search space and the acceptance fraction (Foreman-Mackey et al. 2013). `GALARIO` performs a Fourier transform of the 2D geometric models we generate and creates synthetic visibilities using the observed baseline pairs. The observed Real and Imaginary visibilities are compared to the synthetic equivalents, and the $\chi^2$ and prior probability is minimized[2]. We obtain the residuals by subtracting model visibilities from the observed visibilities and then imaging the difference in `CASA`. This last step allows us to examine the fit quality visually and study features that do not follow our disk model assumptions.

### 2.3. *Statistically assessing the model fits*

We use statistical tests including the Akaike Information Criterion (AIC; Akaike 1974), and the Bayesian Information Criterion (BIC; Schwarz 1978), to qualita-

[2] The minimization is applied on $-\frac{1}{2}\chi^2 + \log p(\theta)$ where $\chi^2 = \sum_{j=1}^{N}\left[\left(\mathrm{Re}_{obs,j} - \mathrm{Re}_{mod,j}\right)^2 + \left(\mathrm{Im}_{obs,j} - \mathrm{Im}_{mod,j}\right)^2\right] w_j$ and $p(\theta)$ is the prior probability for the parameters used in the model. to find the optimal disk parameters.

**Table 3.** Analytic Model Parameters and Priors

| Parameter | Unit | Description | Prior[a] |
|---|---|---|---|
| $F_0$ | mJy | Integrated flux of the PLCT model | $-4 < \log_{10} F_0 < 0$ |
| $\gamma$ | | Surface density gradient | $0.01 < \gamma < 2$ |
| $R_c$ | $''$ | Characteristic disk radius | $0.01 < R_c < 2$ |
| $F_D$ | mJy | Integrated flux of a Gaussian disk model | $-4 < \log_{10} F_D < 0$ |
| $\sigma_D$ | $''$ | $1\sigma$ radius of the Gaussian disk model | $0.01 < \sigma_D < 1$ |
| $\phi$ | | Exponent of the radial drop for the FTG model | $2 < \phi < 10$ |
| $F_R$ | mJy | Integrated flux of a Gaussian ring model | $-4 < \log_{10} F_R < 0$ |
| $\sigma_R$ | $''$ | $1\sigma$ radius of the Gaussian ring model | $0.01 < \sigma_R < 0.5$[b] |
| $\mathrm{loc}_R$ | $''$ | Gaussian ring offset from the center | $0.01 < \mathrm{loc}_R < 1$[b] |
| $F_E$ | mJy | Integrated flux of a Gaussian envelope model | $-4 < \log_{10} F_D < 0$ |
| $\sigma_E$ | $''$ | $1\sigma$ radius of the Gaussian envelope model | $0.5 < \sigma_D < 5$ |
| $i$ | ° | Inclination of the disk | $0 < i < 90$ |
| $P.A.$ | ° | Position Angle of the disk | $0 < P.A. < 180$ |
| ΔR.A. | $''$ | Right Ascension offset from the field center | -[c] |
| ΔDec. | $''$ | Declination offset from the field center | -[c] |

*Notes:*

a The prior ranges are adjusted for particular sources where the literature provides constraints.

b Depending on the source and the structure of the real *uv* visibilities and the imaged residuals of a simpler model, the Gaussian ring and *Gaussian gap* priors are adjusted to focus on particular solutions that yield improved fits of the synthetic visibilities to the observed data.

c ΔR.A. and ΔDec. are measured relative to the phase center. For each disk, we initially fit a 2D Gaussian in the image plane in `CASA` to obtain an initial guess to the offsets and allow Galario to fit for the best offset within a $0.2''$ range of the initial guess.

tively assess the model fits similar to Michel et al. (2022). The AIC is defined as,

$$\mathrm{AIC} = 2k + n \ln \left( \frac{RSS}{n} \right) \tag{4}$$

where $k$ is the number of free parameters, $n$ the number of data points, and $RSS$ is the residual sum of squares. The BIC is evaluated as

$$\mathrm{BIC} = k \ln(n) + n \ln \left( \frac{RSS}{n} \right), \tag{5}$$

which is similar to the AIC except that it has a slightly higher penalty for models with more parameters. For our models $k$ ranges from 6 to 15 while $n{\sim}60000$ and $n{\sim}5000$ for the 1.3 and 0.87 mm observations, respectively[3].

We evaluate ΔAIC and ΔBIC between models to provide a quantitative comparison and determine which model is most statistically favored. Kass & Raftery (1995) quantified $3 < \Delta\mathrm{BIC} < 20$ to be positive evidence in favor of the model with the *lower* BIC value, $20 < \Delta\mathrm{BIC} < 150$ as strong evidence, and $> 150$ as decisive evidence. We use the same comparison scale for ΔAIC. All AIC, BIC results and comparisons are found in Appendix D.

## 3. RESULTS

We classify the disks into three categories: structured (Figure 1), non-structured (Figure 2), and ambiguous disks (Figure 3). Our classification system is based on the *uv* visibility fitting results used to construct an intensity profile rather than examining the image plane intensity distribution. A **non-structured disk** has a smooth *uv* visibility profile that shows no significant deviations, such as a gap or ring **and has negligible residuals**. A **structured disk** requires a Gaussian ring or Gaussian gap feature to fit the disk-scale *uv* visibilities and smooth disk profiles result in symmetrically-structured residuals. An **ambiguous disk** has complex *uv* visibilities and significant residual features such that we cannot rule out the presence or absence of substructure.

[3] The 0.87 mm dataset used for VLa 1623 West has $n{\sim}480000$.

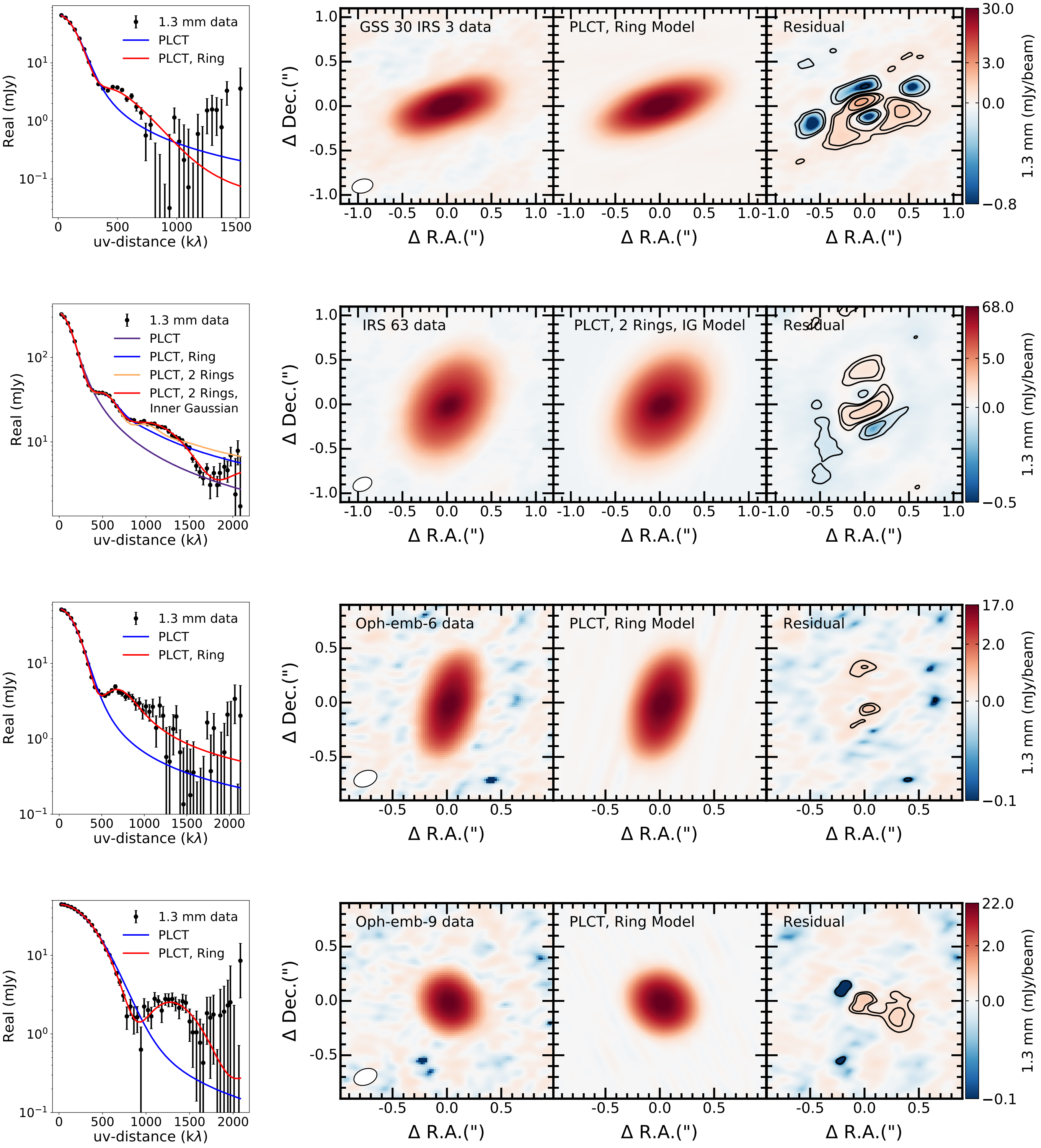

**Figure 1.** Best-fit disk models applied to the 1.3 mm data for the structured disks, GSS 30 IRS 3, IRS 63, Oph-emb-6, and Oph-emb-9. The left column compares the real *uv* visibility data binned in 40 k$\lambda$ bins with different disk models. The statistically favored fit, as evaluated by *AIC* and *BIC* parameters, is always given in red. The three other columns are images of the observed data, the best-fit disk model, and the residuals. All images were made using `tCLEAN` with a *briggs*=0.5 weighting. In the residuals plots, the black contours represent $\pm 3, 5, 10, 15\sigma$ residuals where $\sigma$ is source dependent, see Table 2

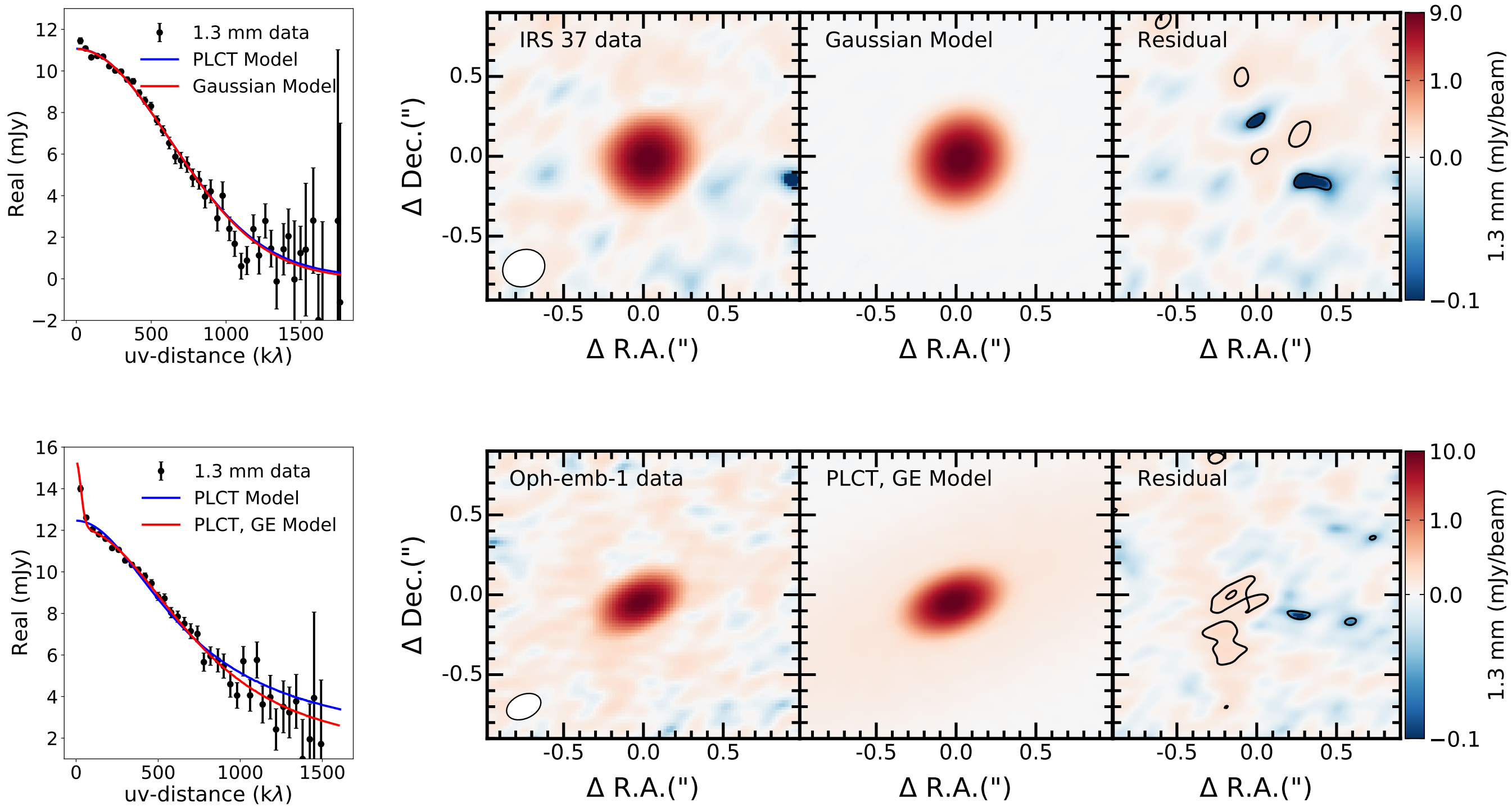

**Figure 2.** Same as Figure 2, but for the non-structured disks, IRS 37 and Oph-emb-1.

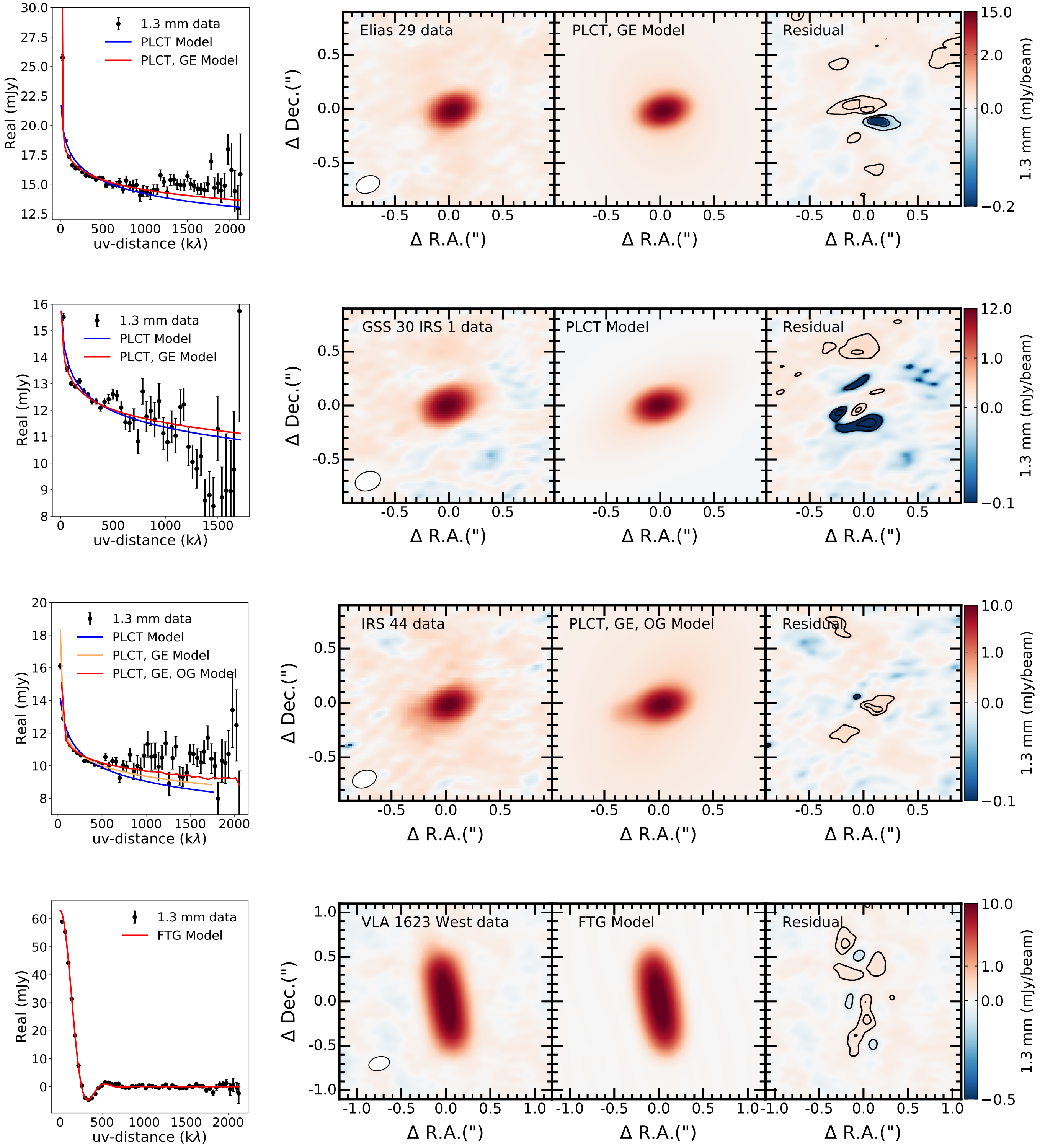

**Figure 3.** Same as Figure 3, but for the ambiguous disks, Elias 29, GSS 30 IRS 1, IRS 44, VLA 1623 West.

In Figures 1, 2, and 3, we compare the best-fit geometric disk models to the observed data at 1.3 mm. The equivalent figures for the 0.87 mm fit results are in Appendix B and generally concur with the 1.3 mm data. For each source, we show the visibility profile of the best-fit disk models over-plotted on the real *uv* data and figures showing the imaged data, models, and residuals. While the visibility profiles show binned *uv*-data, the modelling fits every single *uv* point. Generally, the statistically favored best-fit disk models (see Appendix D) match the sensitive observations with minimal residuals.

GSS 30 IRS 3 (Figure 1, top row) is fit by a PLCT and Gaussian ring disk model. The real *uv* data indicates further scatter and possible structure we do not account for in our best-fit model. Upon examining the residuals, we find these to be up to $+25\sigma_{1.3\mathrm{mm}}$ and $-10\sigma_{1.3\mathrm{mm}}$ and these residuals are similar in morphology to VLA 1623 West at 1.3 and 0.87 mm (see Figure 3 in Michel et al. 2022). The interpretation of the residuals of VLA 1623 West is that it is a flared protostellar disk (Michel et al. 2022). Thus GSS 30 IRS 3 may not only be structured, hosting a ring at 22.8 au but also flared in millimeter emission.

IRS 63 (Figure 1, second row) is the most structured disk in the sample with double Gaussian rings and an inner Gaussian over a PLCT disk at 1.3 mm. The complex disk model fits the data well in the real *uv* plot (left), and the residuals are $\leqslant 15\sigma_{1.3\mathrm{mm}}$. The inner and outer rings are modelled as Gaussians located at 26.5 and 51.0 au from the center with widths of 2 and 13 au, respectively. These are consistent with the higher resolution data from Segura-Cox et al. (2020) (see also, Section 4). The inner Gaussian is a Gaussian function with a small width of 5 au and is only found with the 1.3 mm data. The imaged residuals highlight excess non-axisymmetric emission just south of the disk center spread East-to West in a banana shape. There are also residuals on the northern edge of the disk. While the residuals are at $\leqslant 15\sigma_{1.3\mathrm{mm}}$, they are low-level compared to the disk, representing $\leqslant 1.5\%$ of the peak emission.

Oph-emb-6 (Figure 1, third row) and Oph-emb-9 (Figure 1, bottom row) each have a single Gaussian ring over a PLCT disk model. Both also have low-level residuals, $\leqslant 7\sigma_{1.3\mathrm{mm}}$ and $\leqslant 9\sigma_{1.3\mathrm{mm}}$, respectively. The residuals are also slightly asymmetric and are found within the disks and beyond the disk radii. The rings are located at 32.7 and 21.8 au with 14 and 5 au widths for Oph-emb-6 and Oph-emb-9, respectively.

Figure 2 shows the sample of non-structured disks where the PLCT and Gaussian envelope model is a good fit for Oph-emb-1 while IRS 37 is fit by a simple Gaussian disk. The residuals are very small $\leqslant 5\sigma_{1.3\mathrm{mm}}$, which differentiates these from ambiguous disks where the residuals are $>5\sigma_{1.3\mathrm{mm}}$ and host other irregularities.

Figure 3 shows the sample of ambiguous disks. These have Real *uv* visibility data with significant scatter and deviations from smooth disk profiles resulting in $>5\sigma_{1.3\mathrm{mm}}$ residuals. The best-fit models for Elias 29 (top row) and IRS 44 (third row) account for most of the emission, but the binned real *uv* data are intrinsically noisy and structured at *uv* distances $> 1000$ k$\lambda$. IRS 44 also has an extra offset Gaussian component in its optimal disk model, but this is not considered evidence of the disk's substructures (see Appendix A.2 for further details). For GSS 30 IRS 1, there is noise at $> 500$ k$\lambda$ making the interpretation of the source of *uv* structure ambiguous. VLA 1623 West (bottom row) is complicated to model because the disk is highly inclined at 80° and the emission is likely optically thick. As a result, its *uv* visibilities can be well fit with a structured disk model that includes a gap and ring or a sharp, flat-topped Gaussian model that requires no additional structure, where the latter model is statistically favored (Michel et al. 2022). Since we cannot completely rule out structure in this disk, we consider it to be ambiguous. The residuals of $\leqslant 6\sigma_{1.3\mathrm{mm}}$ are symmetric about the major axis and are also detected at up to $\sim 13\sigma_{0.87\mathrm{mm}}$ at 0.87 mm (see Appendix B), which are reminiscent of a flared disk (Michel et al. 2022).

Table 4 summarizes the resulting best-fit model for each disk. Four protostellar disks (GSS 30 IRS 3, IRS 63, Oph-emb-6, and Oph-emb-9; see Figure 1) are also structured with ring-like features on a PLCT model. Two disks RS 37-A and Oph-emb-1, see Figure 2) are non-structured disks well-fit by a PLCT model. Finally, four protostellar sources (Elias 29, GSS 30 IRS 1, IRS 44, and VLA 1623 West; see Figure 3) are ambiguous disks. For Elias 29 and IRS 44, the disk-scale *uv* visibilities are affected by small- and large-scale emission in the surrounding envelope that can be hard to differentiate from the disk substructure. For VLA 1623 West, the disk is nearly edge-on (80°) and appears flared; we cannot account for the effects from a vertical scale height with our flat disk models (Michel et al. 2022). For the best-fit model intensity profiles, see Appendix C. To confirm that the structured disks are not artefacts from simply being highly inclined, which can appear as a structure in the *uv* visibility space for edge-on protostellar disks (e.g., Michel et al. 2022), we test the FTG model for each structured disk and find that it is systematically disfavored, see Table 7.

In Table 5, we present the best-fit parameters for the optimal model for each protostellar source based on the 1.3 and 0.87 mm data fits. The best-fit numbers rep-resent the median values of the posterior distributions from the last 500 steps of the MCMC chains. The associated uncertainties represent the 68% inclusion interval. We choose the optimal model according to the statistical tests in Section 2.3. The statistical test results for each source are in Appendix D.

**Table 4.** The presence of substructures in the Ophiuchus protostellar disk sample.

| Source | Structured? | Comment |
|---|---|---|
| Elias 29 | Ambiguous | Inner Envelope |
| GSS 30 IRS 1 | Ambiguous | Inner Envelope |
| GSS 30 IRS 3 | ✓ | Gaussian Ring (+ Flaring) |
| IRS 37-A | ✗ | - |
| IRS 44 | Ambiguous | Unknown |
| IRS 63 | ✓ | Two Gaussian Rings (+ Inner Gaussian) |
| Oph-emb-1 | ✗ | - |
| Oph-emb-6 | ✓ | Gaussian Ring |
| Oph-emb-9 | ✓ | Gaussian Ring |
| VLA 1623 West | Ambiguous | Flaring |

**Table 5**. Best-fit parameters for the optimal geometric model.

| | | Best-fit parameters[a] | | | |
|---|---|---|---|---|---|
| Source | Model | $\lambda$ | Unit | 1.3 mm | 0.87 mm |
| Elias 29 | PLCT, Gaussian envelope | $F_0$ | mJy | $21^{+3}_{-3}$ | $45^{+1}_{-1}$ |
| | | $\gamma$ | | $1.98^{+0.02}_{-0.02}$ | $1.94^{+0.48}_{-0.82}$ |
| | | $R_c$ | mas | $257^{+193}_{-155}$ | $289^{+129}_{-190}$ |
| | | $F_E$ | mJy | $33^{+12}_{-13}$ | $81^{+5}_{-26}$ |
| | | $\sigma_E$ | mas | $3120^{+114}_{-245}$ | $2160^{+319}_{-117}$ |
| | | $i$ | ° | $37^{+3}_{-3}$ | $41^{+18}_{-23}$ |
| | | $P.A.$ | ° | $19^{+153}_{-11}$ | $98^{+71}_{-88}$ |
| | | R.A. | h,m,s | 16:27:09.415 | 16:27:09.413 |
| | | Dec. | °,′,″ | -24:37:19.245 | -24:37:19.231 |
| GSS 30 IRS 1 | *PLCT* | $F_0$ | mJy | $16^{+1}_{-1}$ | $50^{+1}_{-1}$ |
| | | $\gamma$ | | $1.81^{+0.02}_{-0.02}$ | $1.59^{+0.02}_{-0.02}$ |
| | | $R_c$ | mas | $462^{+29}_{-71}$ | $492^{+6}_{-11}$ |
| | | $i$ | ° | $66^{+3}_{-2}$ | $57^{+3}_{-3}$ |
| | | $P.A.$ | ° | $119^{+3}_{-3}$ | $103^{+5}_{-4}$ |
| | | R.A. | h,m,s | 16:26:21.356 | 16:26:21.356 |
| | | Dec. | °,′,″ | -24:23:04.915 | -24:23:04.897 |
| GSS 30 IRS 3 | PLCT, Gaussian ring | $F_0$ | mJy | $10^{+4}_{-1}$ | $52^{+10}_{-10}$ |
| | | $\gamma$ | | $0.09^{+0.26}_{-0.06}$ | $0.54^{+0.4}_{-0.4}$ |
| | | $R_c$ | mas | $124^{+12}_{-9}$ | $666^{+31}_{-34}$ |
| | | $F_R$ | mJy | $57^{+1}_{-44}$ | -[b] |
| | | $\sigma_R$ | mas | $218^{+25}_{-30}$ | -[b] |
| | | $\mathrm{loc}_R$ | mas | $164^{+627}_{-27}$ | -[b] |
| | | $i$ | ° | $70^{+1}_{-1}$ | $77^{+1}_{-1}$ |
| | | $P.A.$ | ° | $110^{+1}_{-1}$ | $109^{+1}_{-1}$ |
| | | R.A. | h,m,s | 162621.718 | 16:26:21.718 |
| | | Dec. | °,′,″ | -24:23:50.981 | -24:22:50.959 |
| IRS 37-A | Gaussian disk | $F_D$ | mJy | $11^{+2}_{-2}$ | $25^{+7}_{-7}$ |
| | | $\sigma_D$ | mas | $53^{+1}_{-1}$ | $58^{+2}_{-2}$ |

Continued on next page

**Table 5 – continued from previous page**

| | | Best-fit parameters[a] | | | |
|---|---|---|---|---|---|
| Source | Model | $\lambda$ | Unit | 1.3 mm | 0.87 mm |
| | | $i$ | ° | $70^{+3}_{-2}$ | $72^{+3}_{-3}$ |
| | | $P.A.$ | ° | $9^{+1}_{-1}$ | $7^{+1}_{-1}$ |
| | | R.A. | h,m,s | 16:27:17.581 | 16:27:17.580 |
| | | Dec. | °,′,″ | -24:28:56.833 | -24:28:56.802 |
| IRS 44 | PLCT, Gaussian envelope, Offset Gaussian | $F_0$ | mJy | $14^{+1}_{-1}$ | $34^{+1}_{-1}$ |
| | | $\gamma$ | | $1.85^{+0.04}_{-0.03}$ | $1.60^{+0.05}_{-0.06}$ |
| | | $R_c$ | mas | $366^{+89}_{-126}$ | $232^{+148}_{-100}$ |
| | | $F_E$ | mJy | $10^{+1}_{-1}$ | $37^{+1}_{-1}$ |
| | | $\sigma_E$ | mas | $1366^{+132}_{-110}$ | $1903^{+231}_{-216}$ |
| | | $F_{OG}$ | mJy | $1^{+1}_{-1}$ | $2^{+1}_{-1}$ |
| | | $\sigma_{OG}$ | mas | $5^{+6}_{-3}$ | $26^{+25}_{-17}$ |
| | | $\Delta$R.A.$_{OG}$ | mas | $295^{+6}_{-5}$ | $255^{+16}_{-18}$ |
| | | $\Delta$Dec.$_{OG}$ | mas | $-62^{+4}_{-4}$ | $-32^{+11}_{-13}$ |
| | | $i$ | ° | $61^{+4}_{-4}$ | $59^{+8}_{-7}$ |
| | | $P.A.$ | ° | $147^{+5}_{-5}$ | $169^{+7}_{-7}$ |
| | | R.A. | h,m,s | 16:27:27.987 | 16:27:27.991 |
| | | Dec. | °,′,″ | -24:39:33.954 | -24:39:33.934 |
| IRS 63 | PLCT, Two Gaussian rings Inner Gaussian | $F_0$ | mJy | $237^{+4}_{-4}$ | $575^{+48}_{-96}$ |
| | | $\gamma$ | | $0.57^{+0.01}_{-0.01}$ | $0.71^{+0.03}_{-0.02}$ |
| | | $R_c$ | mas | $302^{+1}_{-1}$ | $318^{+3}_{-4}$ |
| | | $F_{Rin}$ | mJy | $10^{+1}_{-1}$ | $169^{+20}_{-24}$ |
| | | $\sigma_{Rin}$ | mas | $15^{+5}_{-3}$ | $133^{+33}_{-24}$ |
| | | loc$_{Rin}$ | mas | $191^{+2}_{-2}$ | $259^{+23}_{-59}$ |
| | | $F_{Rout}$ | mJy | $69^{+1}_{-1}$ | $91^{+11}_{-7}$ |
| | | $\sigma_{Rout}$ | mas | $91^{+1}_{-2}$ | $77^{+10}_{-7}$ |
| | | loc$_{Rout}$ | mas | $367^{+4}_{-1}$ | $412^{+15}_{-15}$ |
| | | $F_I$ | mJy | $15^{+1}_{-1}$ | -[c] |
| | | $\sigma_I$ | mas | $39^{+1}_{-1}$ | -[c] |
| | | $i$ | ° | $48^{+1}_{-1}$ | $48^{+1}_{-1}$ |
| | | $P.A.$ | ° | $150^{+1}_{-1}$ | $150^{+1}_{-1}$ |
| | | R.A. | h,m,s | 16:31:35.657 | 16:31:35.656 |
| | | Dec. | °,′,″ | -24:01:29.941 | -24:01:29.898 |
| Oph-emb-1 | PLCT, Gaussian envelope | $F_0$ | mJy | $12^{+2}_{-2}$ | $34^{+6}_{-6}$ |
| | | $\gamma$ | | $0.56^{+0.07}_{-0.08}$ | $0.26^{+0.13}_{-0.14}$ |
| | | $R_c$ | mas | $72^{+2}_{-2}$ | $78^{+2}_{-2}$ |
| | | $F_E$ | mJy | $3^{+1}_{-1}$ | $36^{+1}_{-1}$ |
| | | $\sigma_E$ | mas | $1162^{+148}_{-147}$ | $2120^{+221}_{-276}$ |
| | | $i$ | ° | $72^{+1}_{-1}$ | $65^{+1}_{-1}$ |
| | | $P.A.$ | ° | $114^{+1}_{-1}$ | $115^{+1}_{-1}$ |
| | | R.A. | h,m,s | 16:28:21.620 | 16:28:21.616 |
| | | Dec. | °,′,″ | -24:36:24.197 | -24:36:24.166 |
| Oph-emb-6 | PLCT, Gaussian ring | $F_0$ | mJy | $17^{+1}_{-1}$ | $82^{+1}_{-1}$ |
| | | $\gamma$ | | $0.27^{+0.07}_{-0.05}$ | $0.40^{+0.06}_{-0.08}$ |
| | | $R_c$ | mas | $111^{+19}_{-8}$ | $227^{+30}_{-40}$ |
| | | $F_R$ | mJy | $36^{+1}_{-1}$ | $38^{+3}_{-4}$ |
| | | $\sigma_R$ | mas | $104^{+4}_{-6}$ | $89^{+10}_{-10}$ |

Continued on next page

**Table 5 – continued from previous page**

| Source | Model | $\lambda$ | Unit | Best-fit parameters[a] 1.3 mm | 0.87 mm |
|---|---|---|---|---|---|
| | | loc$_R$ | mas | $228^{+17}_{-9}$ | $299^{+20}_{-20}$ |
| | | $i$ | ° | $76^{+1}_{-1}$ | $75^{+1}_{-1}$ |
| | | $P.A.$ | ° | $169^{+1}_{-1}$ | $169^{+1}_{-1}$ |
| | | R.A. | h,m,s | 16:27:05.250 | 16:27:05.251 |
| | | Dec. | °,′,″ | -24:36:30.168 | -24:36:30.153 |
| Oph-emb-9 | PLCT, Gaussian ring | $F_0$ | mJy | $30^{+1}_{-1}$ | $63^{+5}_{-7}$ |
| | | $\gamma$ | | $0.20^{+0.06}_{-0.04}$ | $0.36^{+0.16}_{-0.11}$ |
| | | $R_c$ | mas | $109^{+8}_{-9}$ | $114^{+13}_{-20}$ |
| | | $F_R$ | mJy | $15^{+1}_{-1}$ | $37^{+4}_{-6}$ |
| | | $\sigma_R$ | mas | $38^{+5}_{-8}$ | $42^{+12}_{-17}$ |
| | | loc$_R$ | mas | $157^{+7}_{-7}$ | $154^{+18}_{-15}$ |
| | | $i$ | ° | $67^{+1}_{-1}$ | $65^{+1}_{-1}$ |
| | | $P.A.$ | ° | $28^{+1}_{-1}$ | $27^{+1}_{-1}$ |
| | | R.A. | h,m,s | 16:26:25.473 | 16:26:25.474 |
| | | Dec. | °,′,″ | -24:23:01.852 | -24:23:01.820 |
| VLA 1623 West | FTG | $F_0$ | mJy | $63^{+1}_{-1}$ | $110^{+1}_{-1}$ |
| | | $\sigma_D$ | mas | $449^{+1}_{-1}$ | $459^{+1}_{-1}$ |
| | | $\phi$ | | $5.01^{+0.07}_{-0.07}$ | $4.95^{+0.03}_{-0.03}$ |
| | | $i$ | ° | $80^{+1}_{-1}$ | $81^{+1}_{-1}$ |
| | | $P.A.$ | ° | $10^{+1}_{-1}$ | $10^{+1}_{-1}$ |
| | | R.A. | h,m,s | 16:26:25.632 | 16:26:25.632 |
| | | Dec. | °,′,″ | -24:24:29.618 | -24:24:29.587 |

[a] Errors for all parameters represent the 68% inclusion interval. Errors for R.A. and Dec. position are negligible (typically less than 1 mas) and are excluded from the table.

[b] The ALMA observations of GSS 30 IRS 3 are centered on its companion source GSS 30 IRS 1, such that IRS 3 is significantly offset from the primary beam center. GSS 30 IRS 3 is at 40-43% and 10-12% of the primary beam center at 1.3 mm and 0.87 mm, respectively. The residual images for GSS 30 IRS 3 in Figures 1 are primary beam corrected; the values in this table are not, due to the complexity in addressing the primary beam correction within `GALARIO`.

[c] The ALMA 1.3 mm observations of IRS 63 are more sensitive than the 0.87 mm snapshot observations used in this study. Thus, the 0.87 mm data do not have sufficient sensitivity to model an inner Gaussian.

## 4. DISCUSSION

We find four structured disks (all Class I), two non-structured disks (one Class 0 and one Class I), and four ambiguous disks (all Class I disks), see Table 4. We identify substructures in Oph-emb-6, Oph-emb-9, and GSS 30 IRS 3 for the first time. IRS 63 was previously identified to have structure from higher resolution data (Segura-Cox et al. 2020) and we recover the locations of the two rings that were previously found even though our data have lower resolution. Moreover, we have a consistent ring width as Segura-Cox et al. (2020) for the outer ring, but our inner ring is thinner due to our model including an inner Gaussian which is not in the Segura-Cox et al. (2020) model of IRS 63. They fit the two rings from residual data after subtracting a radiative transfer best-fit smooth disk model. They do not include an inner Gaussian component in fitting the residuals. However, an inner feature can be seen in their residual profile (see their Figure 3). The consistency of the ring locations and the outer ring width highlight the power of $uv$-visibility fitting of highly sensitive data at sub-beam scales and provide a benchmark for the quality of our fits with coarser (35 au) resolution observations.

From our results, 4 out of 10 (40%) of the disks analyzed show substructure, and 30% if we exclude GSS 30 IRS 3, see Appendix A.1. In the context of all 25 identified Ophiuchus protostellar sources, we conclude that at least 4/25 (16%), or 12% without GSS 30 IRS 3, of protostellar disks host substructure, suggesting that substructures in protostellar disks may be as common as at the protoplanetary stage, even though we have small number statistics. In comparison, the frequency of structured disks (rings, transition, and extended) identified in the protoplanetary stage around low-mass ($0.1 - 1\ M_{\odot}$ stars) is 14% (van der Marel & Mulders 2021). The similar occurrence rate for substructures in protostellar and protoplanetary disks could imply that substructure-forming mechanisms in the younger stage are responsible for the observed protoplanetary disk substructures. The possible link between the substructures observed in the protostellar and protoplanetary disks necessitates an underlying substructure-inducing mechanism.

### 4.1. *Formation of substructure.*

Theoretical studies have proposed various mechanisms for forming substructures in circumstellar disks, e.g., zonal flows, envelope infall, dead zones, snow lines, gravitational instabilities, and planets (Andrews 2020). More research is required to identify which process(es) are likely to occur in protostellar disks, but we summarily comment on the various mechanisms' significance in the protostellar disk phases.

*Zonal flows* are caused when magnetic fields threading through the disk repel ionized gas from regions of peak magnetic stress to create pressure bumps (Uribe et al. 2011; Suriano et al. 2018). The pressure bumps halt radial drift to form rings, and the magnetized flows create adjacent gaps. This mechanism has been mainly theorized and studied for protoplanetary disks and speculated to operate in protostellar disks (Suriano et al. 2018).

*Envelope infall* onto the protostellar disk is expected to be periodic and asymmetric via accretion flows. Where envelope material is deposited onto the disk, vortices from Rossby wave instabilities (RWI) can form (Lovelace et al. 1999; Bae et al. 2015). These vortices induce azimuthal shear and produce pressure bumps, which can halt radial drift of millimeter dust. This theoretical mechanism has been simulated for protostellar disks and successfully formed synthetic millimeter dust rings (Kuznetsova et al. 2022).

*Dead zones* are the products of a turbulent/laminar boundary between the inner and outer disks where the density gradient leads to decreasing shielding from UV photons (Dzyurkevich et al. 2013). Vortices can form at the edges of the dead zone through RWI (Lovelace et al. 1999; Lyra et al. 2009). Protostellar disks are expected to be conducive to forming dead zones because they are dense and have high optical depths (see Section 4.4), which shields the inner disk from photo-ionizing radiation, but most work has focused on protoplanetary disks.

*Snow lines* can form annular substructures at varying chemical species sublimation boundaries. While snow lines exist in protostellar disks, they are difficult to identify observationally, and estimates derived from stellar parameters have not yet found a correlation between their positions and known gaps and rings (van der Marel et al. 2019). Future analyses that use chemical tracers from line observations combined with chemical models (e.g., Bianchi et al. 2020; Codella et al. 2021) will be needed to build more comprehensive models of the structured disks and determine the locations of their snow lines relative to the density enhancements.

*Gravitational Instabilities (GI)* occur in disks that are so massive that their own self-gravity causes instabilities and spiral structures (e.g., Lee et al. 2020; Aso et al. 2021). Since protostellar disks are more massive than protoplanetary disks (e.g., Tychoniec et al. 2020; Drazkowska et al. 2022), making them more susceptible to undergoing GI. Spiral substructures have been observed in a few protoplanetary disks at millimeter wavelengths (e.g., Pérez et al. 2016; Rosotti et al. 2020), Only a few studies have found spiral substructures in protostellar disks, TMC-1A (Xu et al. 2023), HH 111 VLA1 (Lee et al. 2020), and a tentative detection in L1527 (Nakatani et al. 2020; Ohashi et al. 2022; Sheehan et al. 2022b), but the latter is difficult to confirm due to a high inclination ($i > 85°$). Of the 10 disks in our sample, only IRS 63 has an estimated stellar mass necessary to calculate disk-to-star mass ratios and determine the stability of the disk against self-induced GI. Adopting a stellar mass for IRS 63 of 1 $M_{\odot}$[4] (Segura-Cox et al. 2020) and a disk mass of 50 $M_{Jup}$ (Sadavoy et al. 2019), the disk-to-star mass ratio is 0.05, indicating it may be unstable to GI.

*Planets* can also form disk substructures through gravitational perturbations and accretion (Lin & Papaloizou 1979; Goldreich & Tremaine 1980; Paardekooper et al. 2022). A planet can (1) clear a dust lane which acts as a barrier for dust grains drifting inwards and results in the formation of a ring outside of the newly opened gap (Zhu et al. 2012; Pinilla et al. 2012, 2020); (2) induce planet-driven spiral waves (Ogilvie & Lubow 2002; Bae & Zhu 2018; Speedie et al. 2022); (3) perturb the disk to create multiple gaps and rings (Dong et al. 2017, 2018). A few protoplanetary disks have had planets and kinematic signatures of planets detected within cavities or gaps in their disk in support of this planet-induced substructure, e.g., PDS 70 (Keppler et al. 2018; Müller et al. 2018; Haffert et al. 2019); HD 163296 (Pinte et al. 2018); HD 97048 (Pinte et al. 2019); AB Aurigae (Currie et al. 2022); AS 209 (Bae et al. 2022). To carve out gaps and cavities, planets need to grow to sufficient mass, and they may not have enough time to reach those masses in the protostellar phase, <0.5 Myr (Evans et al. 2009; Dunham et al. 2014b, 2015). The formation of planets from the growth of dust grains on such short time scales is challenging for models (e.g., Lambrechts et al. 2019; Raymond & Morbidelli 2022) and suggests that planets are unlikely to produce substructures in protostellar disks unless the process is very rapid as explored in Lee et al. (2022). Detecting larger bodies such as planetesimals (kilometer-sized bodies) or planetary cores in a protostellar disk is a challenge that could be tackled through high-resolution line observations. Future line observations will be necessary to look for planetary wakes and other disturbances that cause deviations from Keplerian rotation (Perez et al. 2015; Calcino et al. 2022).

[4] IRS 63's mass is estimated from models of protostars evolving from a collapsing core and is thus not directly attributed to observational results.

### 4.2. *Which protostellar disks have detectable substructure?*

Figure 4 shows the model integrated flux of each protostellar disk grouped by type. For sources that require a Gaussian envelope, we did not include this component in the integrated flux evaluation, given that it does not pertain to the protostellar disk[5]. We find a trend between the detection of substructures and disk brightness. The structured disks tend to be bright, whereas the non-structured disks tend to be faint. The ambiguous disks are in the middle. This trend is consistent across the 1.3 and 0.87 mm data. If we use the integrated flux as a proxy for mass[6], assuming that the dust emission is optically thin (Hildebrand 1983), we would infer that the structured disks are also those with higher masses. A similar trend is also seen in the Class II disk study by van der Marel & Mulders (2021), where structured disks are found to be more massive and non-structured (compact) disks have lower dust masses. The similarity suggests that substructures are already created in the embedded Class I stage.

Figure 5 compares the disk size for each protostellar source in the same fashion as Figure 4. We evaluate the disk size as the radius that contains 90% of the integrated flux from the protostellar disk's best-fit model[7] (e.g., similar to Cieza et al. 2021). We use $R_{90\%}$ since the PLCT model includes an inner envelope component expected to have a low flux contribution relative to the disk. We do not use $R_c$ because the transition from the disk to the inner envelope is unclear when $\gamma$ is large. The uncertainty associated with $R_{90\%}$ corresponds to the standard deviation from the resulting distribution of $R_{90\%}$ values from bootstrap sampling the disk parameters within their uncertainty ranges we describe as normal distributions. We find that the structured disks tend to be larger with a median $\tilde{R}_{90\%} = 37$ au compared to the non-structured disks where $\tilde{R}_{90\%} = 14$ au. The ambiguous disks have $\tilde{R}_{90\%} = 34$ au, comparable to the structured disks. For the protostellar disks we find a bias towards larger disks hosting substructures compared to smaller ones.

While we suggest that the non-structured disks do not host significant disk substructure, we cannot exclude the possibility of small or faint substructures below our detectability threshold. For example, very high resolution and high sensitivity observations of compact, low-mass protoplanetary disks have found physically small substructures where the disks were previously considered smooth (Long et al. 2018; Kurtovic et al. 2021). Never-theless, it is unlikely that these disks will become large-ringed protoplanetary disks (e.g., Andrews et al. 2018) or host deep gaps as found in transition disks (e.g., Francis & van der Marel 2020) due to the correlation between the mass of the protoplanetary disk and the strength of dust traps halting radial drift. Pinilla et al. (2020) find that lower mass stars likely host lower mass disks that will form weaker dust traps and thus cannot produce large-scale rings nor deep extended cavities within their disks.

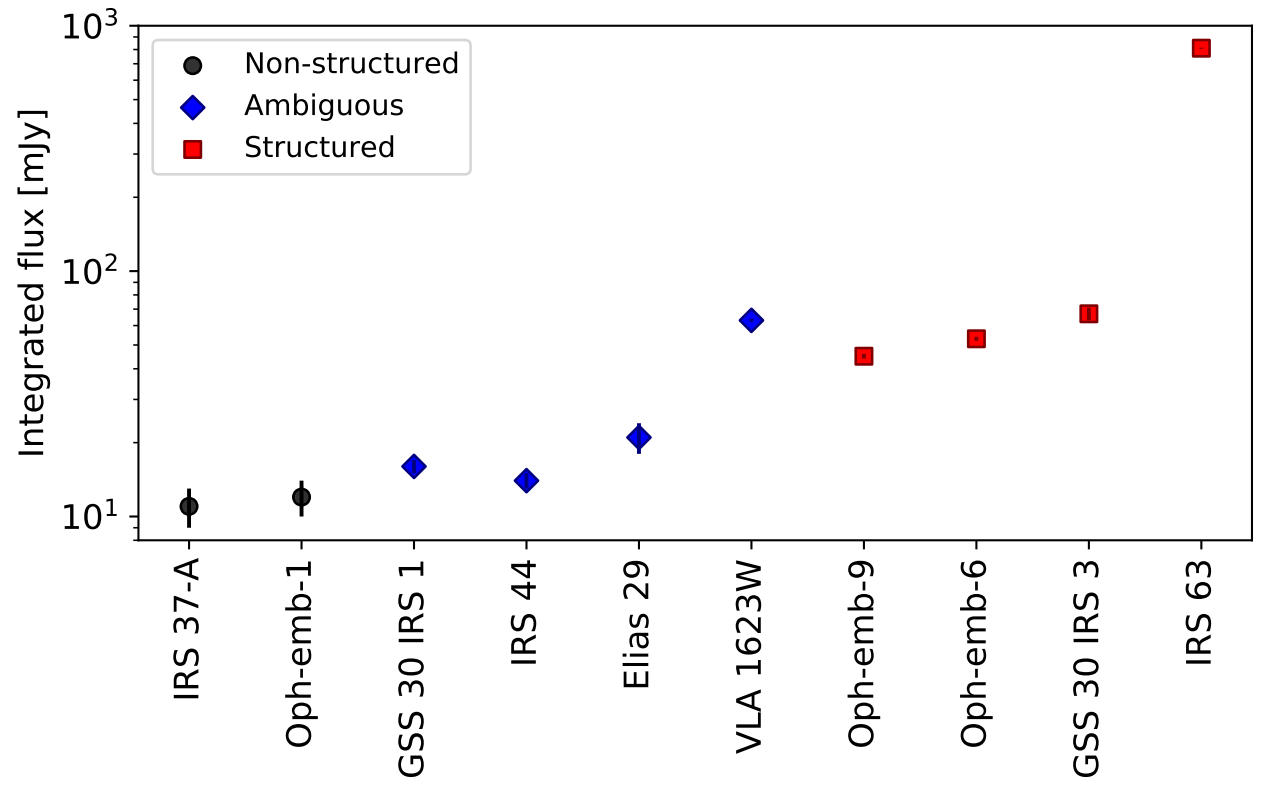

**Figure 4.** 1.3 mm integrated flux per disk type evaluated from the best-fit protostellar disk model. Sources are grouped according to their disk substructure. The integrated flux and corresponding uncertainties shown as error bars are in Table 5. The uncertainties calculated represent the 68% inclusion interval from the MCMC runs, see Section 3, and do not include flux calibration uncertainties.

[5] The PLCT model fits both the disk and the inner envelope, so the integrated flux does include some envelope emission. However, this contribution is small as an exponential tail models the envelope compared to a power law for the disk

[6] Given the limitations surrounding this calculation, we do not make the flux conversion to mass for the protostellar disks due to the unknown dust opacities, dust-to-gas ratios, disk temperature structures, and optical depth corrections.

[7] In our calculations for the disk size that contains 90% of the flux, we exclude the inner 24 and 13 milliarcseconds at 1.3 and 0.87 mm, respectively. This corresponds to one-tenth of the beam at each wavelength and mitigates the effect of the small grid size with respect to the analytic power-law fits, which disproportionally include significant amounts of flux at radii close to zero.

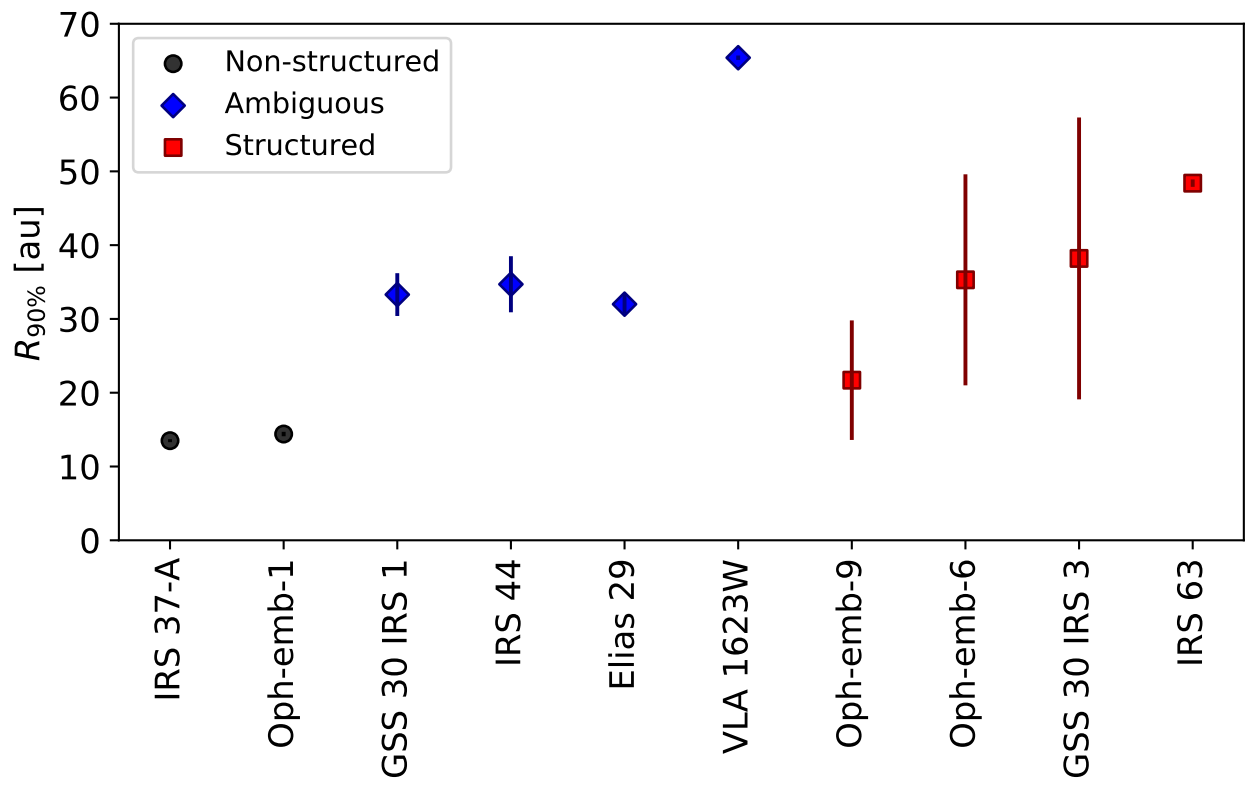

**Figure 5.** Same as Figure 4 for the protostellar disk size evaluated as $R_{90\%}$ based on the 1.3 mm best-fit disk models. GSS 30 IRS 3 is found to have large uncertainties on the disk size owing to the uncertainties on the location of the ring feature thus propagating into the error calculations for $R_{90\%}$, we thus artificially limit the uncertainty to 50%.

### 4.3. *What are the substructures properties?*

#### 4.3.1. *No evidence of gaps or spirals.*

We identify substructures in emission based on deviations from a smooth disk model. We only find enhancements above a smooth disk profile among the structured disk sample. We do not find gaps or deficits in emission. The absence of gaps is important in describing the origin of disk substructures. Whichever mechanism forms substructures at the protostellar stage appears only to produce enhancements. This result disfavors large Jupiter-mass planets, zonal flows, dead zones, and snow lines as the origin of the rings, as these mechanisms usually produce a gap followed by a ring(s). The absence of resolved gaps suggests that they are either below our detection limits or that we need an alternative mechanism that does not produce a gap to explain the enhancements seen in this sample of protostellar disks.

In addition to seeing no evidence of gaps, we find no signatures of spiral substructures, see Appendix A.1. Protostellar disks tend to be more massive than protoplanetary disks (Kratter & Lodato 2016; Tobin et al. 2020), and as such, we could expect GI-induced spirals. Nevertheless, we do not observe spiral substructures in our protostellar disk sample. We may be insensitive to spiral substructures with the current observations, which may be hidden due to the optical depths. For example, spiral waves can exist in the smaller dust or the gas above the mid-plane (e.g., HD 34282; de Boer et al. 2021; Benisty et al. 2022) and would not be detected in the ALMA continuum observations at 1.3 or 0.87 mm. Another reason may be that the spiral features are below the observed resolution, and higher resolution data is needed to rule this out firmly.

#### 4.3.2. *The protostellar ring features.*

We describe the ring substructures using Gaussians (as described in Section 2.2). Figure 6 compares ring location ($\mathrm{loc}_R$) and the ring width ($\sigma_R$), normalized by the disk size as given by $R_{90\%}$. The ring locations vary from the outer edges in the cases of Oph-emb-6, Oph-emb-9, and IRS 63 (outer ring) to midway across the disk for GSS 30 IRS 3 and IRS 63 (inner ring).

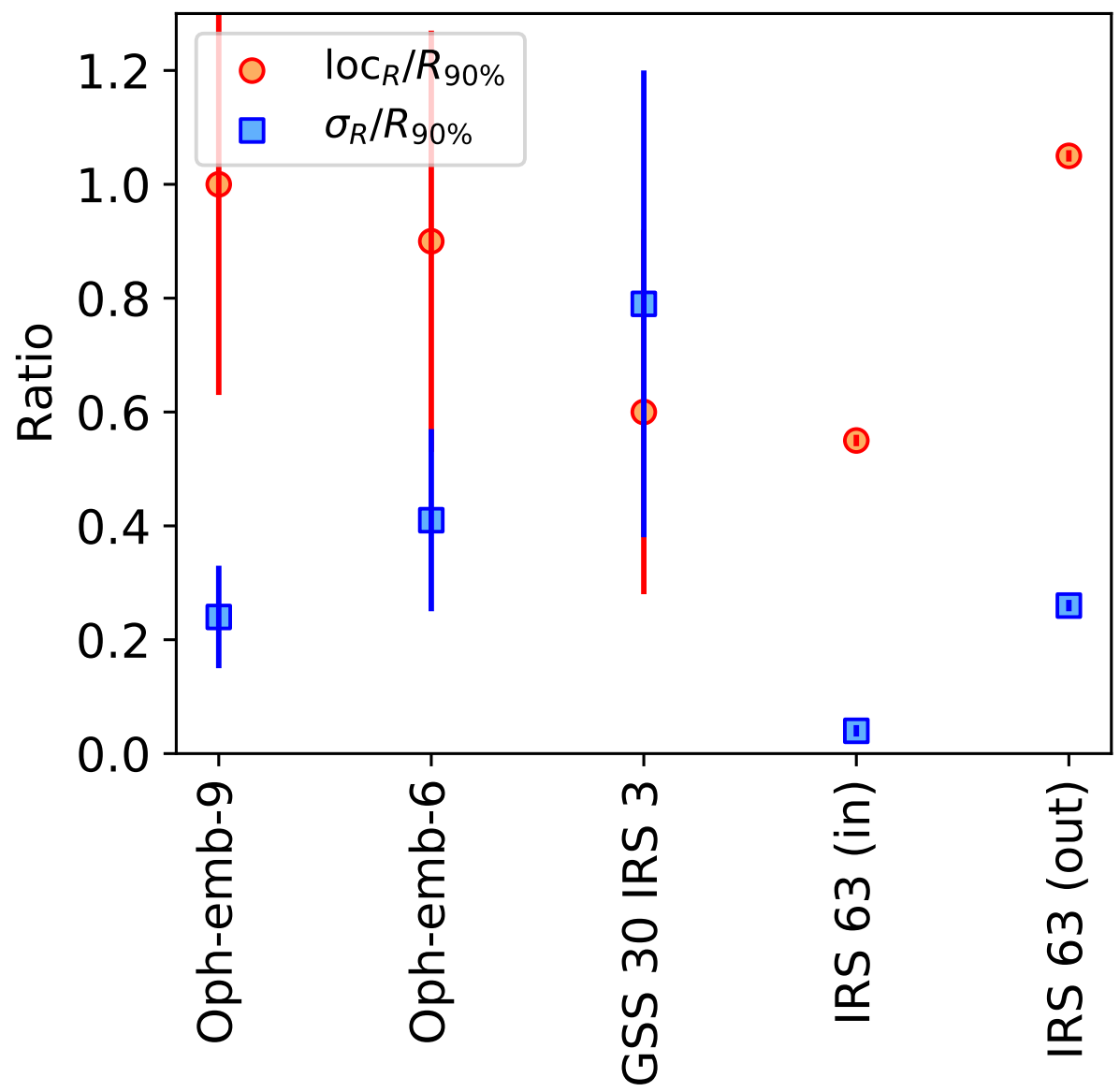

**Figure 6.** The ratio of the Gaussian ring location ($\mathrm{loc}_R$) to the disk size ($R_{90\%}$) is shown in orange and the ratio Gaussian ring width ($\sigma_R$) to the disk size ($R_{90\%}$) is in blue. Both are based on the 1.3 mm best-fit disk models and parameters in Table 5. The error bars displayed are the quadrature sum of the errors associated with $\sigma_R$ and $R_{90\%}$ to represent the general uncertainty.

For $\sigma_R/R_{90\%}$, we find a median of 0.26, suggesting that the rings are typically a quarter of the disk's radius. This broad median width relative to disk size (see Section 4.6) is interesting as it may be a feature of the physical origin of the rings themselves. Alterna-tively, the resolution could limit our ability to constrain the Gaussian ring widths precisely. While *uv*-visibility fitting allows us to probe at sub-beam scales, we are intrinsically limited by the observed *uv* sampling. Nevertheless, we obtain reliable results as evidenced with IRS 63, where we find a best-fit model consistent with higher beam resolution observations from Segura-Cox et al. (2020).

One source, GSS 30 IRS 3, has $\sigma_R > \text{loc}_R$ (see Figure 6 and Table 5), which we attribute to challenges in constraining its model due to possible flaring (see Section 3). Flaring has been seen at millimeter wavelengths in other protostellar disks (e.g., HH 212, l1527; Lee et al. 2017; Ohashi et al. 2022; Sheehan et al. 2022b; Michel et al. 2022) and can complicate the interpretation of the *uv* visibilities. Higher resolution data or data at longer wavelengths (where flaring is less significant) will help constrain the structure of this disk.

### 4.4. *Protostellar inner envelopes*

The PLCT model has two components: the power law and the exponential tail, which are connected by the surface density gradient parameter, $\gamma$. The protostellar sources we fit with the PLCT model are embedded objects, meaning there are disk and envelope components. We detect envelope emission, although the largest scales are filtered out. The exponent, $\gamma$, ranges from 0 to 2, where $\gamma \to 2$ means a smooth, gradual transition between the envelope and disk, and the envelope contribution is important at the disk edge. Lower values of gamma indicate a sharper transition. When $\gamma \to 0$, there is no envelope, and the emission is modelled well by a Gaussian-type profile.

We included an extra Gaussian envelope component for some sources because the PLCT model alone does not fully capture the most extended emission. We note that this terminology diverges from typical nomenclature where classically, the protostellar inner-envelope corresponds to scales $R \leq 3000$ AU (van Dishoeck 2006; Miotello et al. 2014). Our biggest large-scale Gaussian envelope has a $\sigma_E$ of 430 au at 1.3 mm and would classically still be part of the inner envelope. Nevertheless, this extra Gaussian envelope component corresponds to a more extended structure than what is captured by the PLCT model, and we make this distinction between the two "envelopes". Hereafter, we focus on the inner envelope measured by the PLCT model and the $\gamma$ exponent.

Figure 7 shows the 1.3 mm and 0.87 mm best-fit PLCT model $\gamma$ values as a function of the protostellar disks grouped by type. The $\gamma$ values evaluated at the different wavelengths are consistent. All four structured disks have low but non-zero $\gamma$ values. The ambiguous disks, excluding VLA 1623 West, have very high $\gamma$ values. VLA 1623 West was previously thought to have close to no envelope (Murillo et al. 2013; Kirk et al. 2017; Michel et al. 2022). The edge-on nature of VLA 1623 West complicates the description of an envelope, but recent results from Mercimek et al. (2023) estimate it to be 0.04 $M_\odot$, and they also find accretion streamers adding material onto the protostar and disk system. In addition, the ambiguous disks IRS 44 and Elias 29 each need the extra Gaussian envelope component. The non-structured disks have a small spread in $\gamma$ values. IRS 37-A is better fit by a Gaussian disk than a PLCT model, implying no envelope is traced by the millimeter emission ($\gamma \to 0$), whereas Oph-emb-1 is fit with a medium $\gamma$ value and a Gaussian envelope.

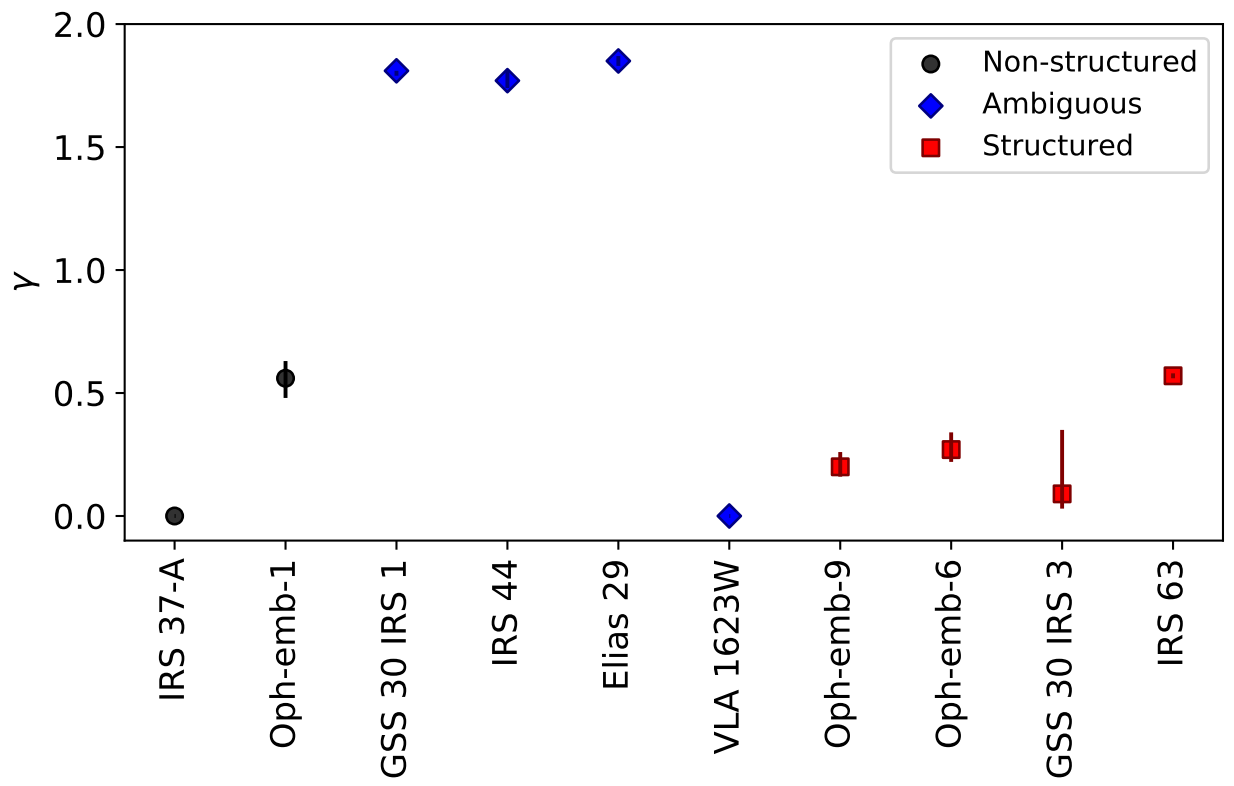

**Figure 7.** Same as Figure. 4 for the protostellar source surface density gradient per disk type from the 1.3 mm best-fit PLCT models. IRS 37-A and VLA 1623 West are fit with a Gaussian disk and a *Flat-Topped Gaussian* model, respectively, since the best-fit PLCT $\gamma$ parameter goes to zero for these two sources. The 0.87 mm data show consistent results.

During protostellar evolution, envelope mass is expected to decrease ($\gamma$ should decrease) with time (Andersen et al. 2019; Sheehan et al. 2020). In comparison, Andrews et al. (2009) fit the same PLCT model to a sample of protoplanetary disks in Taurus and found only $\gamma < 1$. Thus, the structured disks may be more evolved protostellar objects with low $\gamma$ values. Nevertheless, Oph-emb-1 and VLA 1623W are identified as the sample's youngest objects and each have relatively low $\gamma$ values. This contradiction suggests that the link between $\gamma$ and evolution may be weak. The $\gamma$ value from the PLCT model may not fully capture the presence or absence of an envelope, particularly due to filtering out large scales and it being probed by dust emission. Both sources may also be non-standard. Oph-emb-1 has been classified as a proto-brown dwarf and may have a low envelope mass (Hsieh et al. 2019). VLA 1623W is a complicated source that may have been ejected from a higher-order multiple system, thereby losing its envelope (Harris et al. 2018), or it could be non-coeval (Murillo et al. 2016). Nevertheless, we find three ambiguous disks with very high $\gamma$ values, possibly hinting that these objects where there is confusion between envelope and disk material are deeply embedded behind a thick structured envelope shell.

### 4.5. *Comparison to Orion protostellar disks*

We compare our sample of structured protostellar disks to the structured disks identified in Orion, as Orion is the only other cloud with a uniform multi-target sample of protostellar disks with substructure to date. While structured disks have been found in individual sources in nearby clouds, these observations are piecemeal and non-uniform, making comparisons to Ophiuchus more challenging. In Orion, Sheehan et al. (2020) identified seven isolated protostellar disks out of a sample of 328 protostars with confirmed substructure. As a first look at the different structured protostellar disk populations across regions, we compare these Orion disks with our sample of four Ophiuchus structured disks.

All seven Orion protostellar disks in Sheehan et al. (2020) have inner gaps or cavities devoid of dust, akin to transition disks. These substructures differ from what we observe in our Ophiuchus disk sample. Sheehan et al. (2020) do not use an underlying disk model given the cavities. Instead, they use modified Gaussian rings, asymmetric and point source components, and a large-scale Gaussian envelope component.

In Table 6, we compare the ring locations and widths between our sample of protostellar disks and the Orion protostellar disks (Sheehan et al. 2020). The Orion protostellar rings tend to be wider than the Ophiuchus protostellar ring features, which could indicate a different formation mechanism. This result is true whether we consider the full Orion sample or a restricted $\lesssim$ 70 au sample of Orion disks to better match our Ophiuchus population. Sheehan et al. (2020) suggest that various mechanisms can induce the observed substructures found in Orion. They highlight that cavities in such young disks are hard to explain. It could be related to either rapid planet formation or, more likely close-separation binary formation. However, they do not find evidence for multiplicity in their sample.

Unlike the Orion disks, we do not find gaps in the Ophiuchus disks. We note, however, that a comparison between Ophiuchus and Orion protostars is challenging because of the different cloud environments and observational biases. For example, the Orion disks were identified as being structured from visual identification in the image plane rather than *uv* visibilities as was done here. As such, only large substructures like broad rings, were found. Therefore, these are larger disks and substructures and may not be a fair comparison to those found in Ophiuchus. Moreover, the protostars in Orion may be a different population than those protostars in Ophiuchus. Broadly Ophiuchus is expected to form primarily low-mass (M, K) stars, whereas Orion is forming stars of higher mass (Bally 2008). Even the protostars in Orion appear to have higher mass based on simple models (Sheehan et al. 2022a). Stellar mass affects disk properties and evolution (Appelgren et al. 2020; Sellek et al. 2020; Concha-Ramírez et al. 2022). In the protoplanetary disk stage, disk mass and stellar mass are correlated (Pascucci et al. 2016), and so is the frequency of structure and disk mass (van der Marel & Mulders 2021). Therefore any differences between the substructures seen in the Ophiuchus and Orion disks should be taken with caution.

### 4.6. *Comparison to protoplanetary disks*

We compare our disks with the DSHARP sample of structured protoplanetary disks (Huang et al. 2018). The DSHARP sample was selected to obtain high-resolution data from low-mass star-forming regions, which are more representative of the protostars in Ophiuchus. Their disks are all within $\lesssim$ 160 pc, and most are large and massive among the low-mass star-forming region sample (Andrews et al. 2018). While our data were taken from an unbiased survey, we selected the brightest protostellar disks, and thus, our sample shares a similar bias to the DSHARP sample. Table 6 includes the ring location and widths from the DSHARP sample. While the beam resolution of our Ophiuchus protostellar disks is larger than the DSHARP protoplanetary disks, our investigation is in the *uv* plane, different from the image plane for the protoplanetary disks. This allows us to retrieve features at sub-beam scales, at similar physical size scales as the protoplanetary disks. We qualify this statement based on our ability to find IRS 63's substructures at the same physical scales as Segura-Cox et al. (2020), who used 5 au resolution observations, which matches DSHARP.

For the 4 disks that have rings in DSHARP with radii <70 au (matching the Ophiuchus protostellar disks we sampled), the median ring width was 5.4 au (SR4, Elias 20, HD 14266, RU Lup, Sz 114 - excludes upper limits) (Huang et al. 2018). For the structured protostellar Ophiuchus disks, the median is 13 au (regardless of the inclusion of GSS 30 IRS 3). So for the subset of

**Table 6.** A comparison of the ranges of the ring properties between our sample, the Orion protostellar disks, and DSHARP protoplanetary disks. These results are based on ALMA 1.3 mm observations, this sample's protostellar disks, Orion protostellar disks (Tobin et al. 2020), and DSHARP protoplanetary disks (Andrews et al. 2018).

| | Protostellar disks | | Protoplanetary disks |
|---|---|---|---|
| | This sample | Sheehan et al. (2020)[a] | Huang et al. (2018)[b] |
| Number of disks | 4 | 7 | 18 |
| Ring location (in au) | 22-52 | 42-238 | 6-155 |
| Ring width (in au) | 2-30 | 16-116 | 2-20 |
| Median ring width[c] (in au) | 13 | 26 | 5.4 |
| Beam resolution[d] (in au) | 35 | 32 | 5 |
| Sensitivity (in $\mu$Jy bm$^{-1}$) | 30-90 | 330 | 10-20 |

[a] The ring locations and widths are from the analytic model fitting of the sources found in Table 3 in Sheehan et al. (2020). To convert the milliarcsecond results to au, we use a distance of 400 pc for Orion (Kounkel et al. 2017).

[b] The ring locations and widths are obtained by fitting ellipses in the image plane of the sources found in Table 1 of Huang et al. (2018).

Measured only for disks with sizes $\lesssim$ 70 au to match the Ophiuchus sample.

Assuming a robust weighting of 0.5 during imaging.

DSHARP with similar disk sizes, the Ophiuchus rings do appear broader. To compare with the full DSHARP sample, we scale by disk size. Figure 8 compares ring width normalized by disk size between protostellar and protoplanetary disk samples, but this work evaluates disk size differently from Huang et al. (2018). For protoplanetary disks, the disk size is defined fits of ellipses (Huang et al. 2018). A comparison of the ring width to disk size ratio is used to account for the small number statistics and thus the different disk radii ranges and values between the protostellar disks, $\tilde{R}_{90\%} = 37$ au median for a 22 to 48 au range, and protoplanetary disks, $\tilde{R}_{\rm dust} = 85$ au median for a 27 to 264 au range. Most of the protoplanetary disks have narrow rings with a median $\sigma_R/R_{\rm disk}$ of 0.08, while for the structured Ophiuchus and Orion protostellar disks, the rings are much wider with $\sigma_R/R_{\rm disk}$ medians of 0.26 and 0.31, respectively. We could also be limited by the coarser resolution, where it is difficult for the 35 and 32 au resolution observations of the Ophiuchus and Orion protostellar disks, respectively, to identify the sharp and narrow substructures from the *uv* visibilities. However, we find consistent narrow ring widths in IRS 63 similar to the much higher resolution ALMA data from Segura-Cox et al. (2020), which suggests that for the Ophiuchus protostellar disks, the data are sensitive enough to detect small-scale and possibly narrow features robustly.

Figure 9 compares ring location normalized by disk size between our protostellar disks and the DSHARP protoplanetary disks, similar to Figure 6. For the protoplanetary disks, there is a broad range. Some disks, e.g., AS 209, have rings at all radii while other disks, e.g., GW Lup, only have rings at large radii ($\mathrm{loc}_R/R_{\rm disk} > 0.81$).

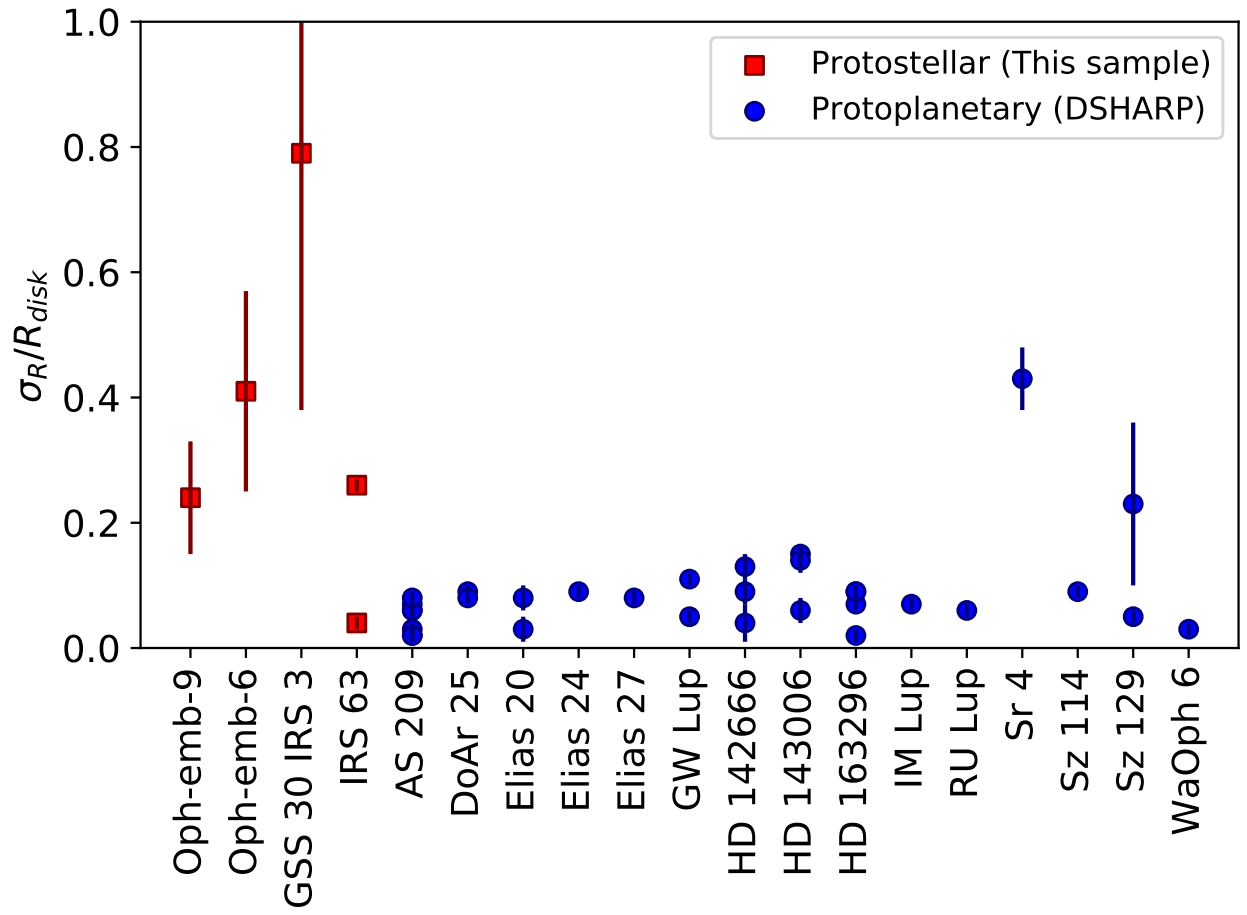

**Figure 8.** The 1.3 mm ring width to disk size ratio of the structured protostellar disk sample (in red) compared to the protoplanetary disks (in blue) observed by DSHARP (Huang et al. 2018).

Ultimately, we find no clear trend in the ring locations in the disks between the protostellar and protoplanetary stages.

We use the same Gaussian functional form for the substructures as protoplanetary disk studies (e.g., Dullemond et al. 2018; Huang et al. 2018; Guzmán et al. 2018; Isella et al. 2018; Pérez et al. 2018; Macías et al. 2019). However, there exists a difference in the fundamental disk properties: we fit Gaussian rings on top of a PLCT disk model. Therefore, the Gaussian rings we identify are flux enhancements above a smooth underlying disk. Unlike many structured protoplanetary disks where the dust emission significantly drops at specific radii, we do not identify such gaps in the protostellar disk emission.

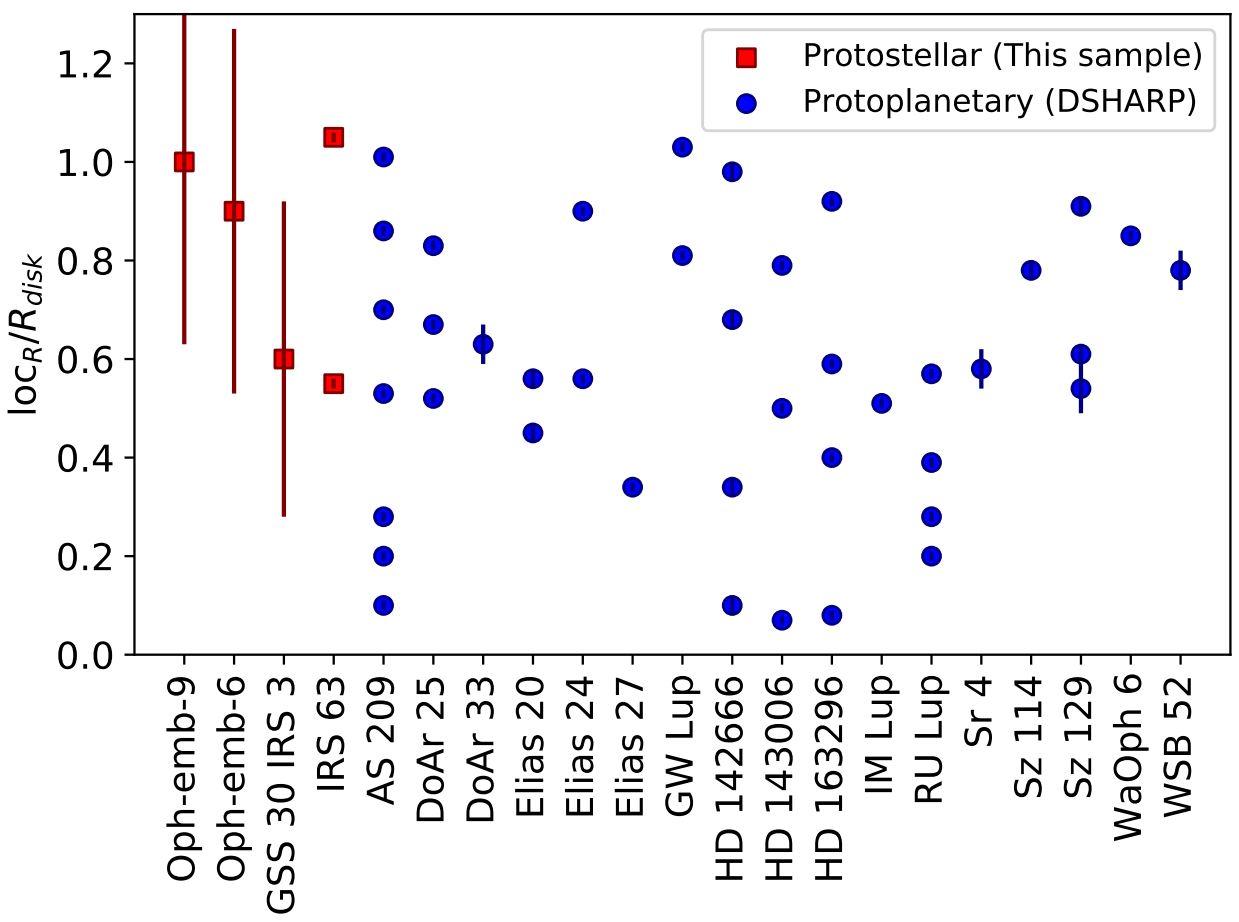

**Figure 9.** The 1.3 mm ring location to disk size ratio of the structured protostellar disk sample (in red) compared to protoplanetary disks (in blue) observed by DSHARP (Huang et al. 2018).

To compare the ring features between our protostellar disks and the DSHARP protoplanetary disks, we use the contrast parameter that has been previously applied to protoplanetary disks. Huang et al. (2018) use intensity profiles from the image plane to identify gap-ring pairs and measure the minimum ($I_{\rm min}$) and maximum ($I_{\rm max}$) depths (see their Figure 1). They define the contrast as $C = I_{\rm min}/I_{\rm max}$. We do not identify gaps or have sufficient resolution to see the rings in the image plane for our sample of protostellar disks. To approximate a similar contrast measure, we use a modification of the gap depth from Huang et al. (2018) and define a ring contrast, $C$, as,

$$C = 1 - \frac{I_{\rm min}}{I_{\rm max}} \approx \frac{I_{\rm res}}{I_{\rm obs}} \qquad (6)$$

where $I_{\rm res}$ is the residual intensity after a smooth disk is subtracted, and $I_{\rm obs}$ is the observed disk brightness. We evaluate the imaged ring residuals by subtracting the best-fit smooth PLCT model from the observations. We select a beam-shaped area at the maxima of the ring residuals and use the mean flux in this area to determine $I_{\rm res}$ and $I_{\rm obs}$. While this is not an exact comparison to the protoplanetary disk $1-I_{\rm min}/I_{\rm max}$ ratio, it provides a comparable measure of relative enhancement of the ring compared to the underlying emission.

Figure 10 compares the contrast ratio between our structured protostellar disks and the protoplanetary disks observed by DSHARP (Huang et al. 2018). Most of our protostellar substructures have very low contrast, whereas the protoplanetary disk substructures show great variety. In the DSHARP disks, there is a range from very high contrasts nearing one and low contrasts approaching zero. The protostellar disks tend to have contrasts $< 0.2$. The exception, GSS 30 IRS 3, is less constrained, likely due to flaring (see Section 3) such that its ring is large (see Table 5) and more of a prominent feature than a minor enhancement like the other cases. The other disks have much lower contrast values, leading us to conclude that these substructures are enhancements rather than stand-alone rings like the protoplanetary disks from Huang et al. (2018). The observed contrast and width ($\sigma_R/R_{\rm disk}$) variations could be evolutionary effects whereby the original substructures are broad relative to disk size and have low contrast and then evolve into narrower features with a mix of contrast ratios.

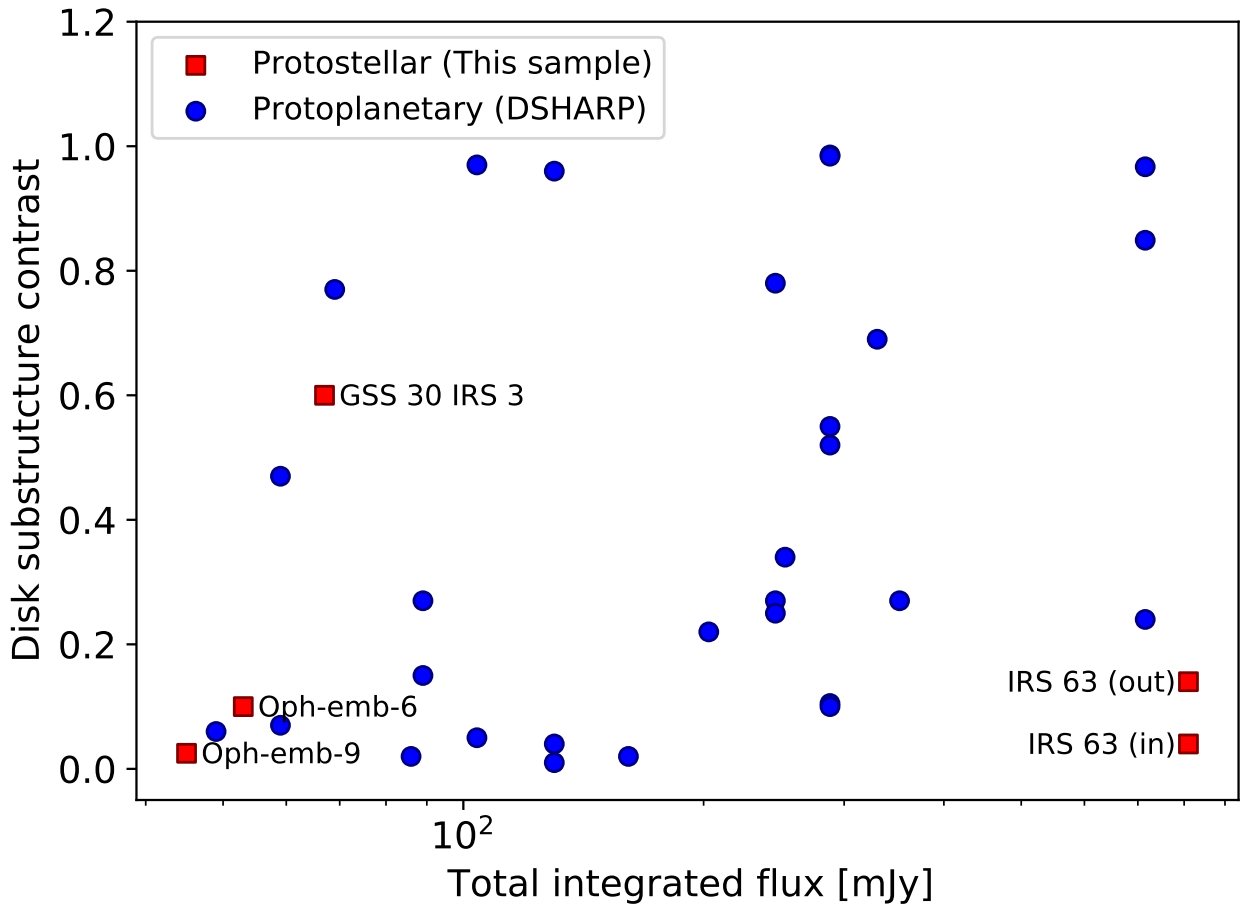

**Figure 10.** The 1.3 mm contrast of the structured disk features from the protostellar disk sample (in blue) compared to protoplanetary disks (in red) observed by DSHARP (Huang et al. 2018). IRS 63 and several protoplanetary disks have multiple data points, given the presence of two or more ring features identified in the disks. Given that the two methods to evaluate this parameter differ, we do not plot the error bars for the contrast. Given a 10% uncertainty for ALMA, fluxes are generally expected and consequently conservatively estimate that errors for the contrast should be $< 20\%$, as for the protoplanetary disks, a 10% uncertainty is generally reported by Huang et al. (2018).

### 4.7. *Substructure evolution and implications for planet formation*

The observed rings in our Ophiuchus sample appear as broad relative to disk size and shallow enhancements above smooth disk profiles. How these shallow density enhancements were first formed remains unclear. Since we do not observe gaps or spiral features in the disks, most of the main mechanisms behind substructure formation in disks do not fully explain these observations. Nevertheless, most of the mechanisms have mostly been explored in protoplanetary disks, which are gas-poor, optically thin, and non-accreting from an envelope. A mechanism most suitable for the conditions of a protostellar disk is structure formation via envelope infall where rings are produced from RWI (Bae et al. 2015; Kuznetsova et al. 2022). The biggest divergences between our results and Kuznetsova et al. (2022) are that they find gaps and the width of these ring features. We only find broad and shallow dust rings, while the theoretically simulated rings from Kuznetsova et al. (2022) are narrower, on the order of half of the scale height of the disk at the ring location.

How the initial rings observed in these structured protostellar disks evolve after formation has yet to be explained. However, we find that the brightest disks are more likely to be structured (Figure 4), in agreement with correlations seen in protoplanetary disks. The most massive protoplanetary disks are statistically more likely to host large-scale disk substructures than the lower mass disks, which are compact and do not display significant substructures (van der Marel & Mulders 2021). The van der Marel & Mulders (2021) protoplanetary disk sample is quasi-complete for all nearby regions within 300 pc. In contrast, the protostellar disk sample analyzed in this work is limited to the Ophiuchus region. It represents 10/25 protostellar sources, of which the ten selected sources are the most massive single or wide binaries. Nevertheless, the agreement between the two samples suggests that structured protoplanetary disks may evolve from structured protostellar disks. The ambiguous disks are hard to characterize and may also be the precursors to structured protoplanetary disks, but further study is needed.

The possible connection between these two disk populations can be explained through two pathways: (1) the protostellar rings will form planets at those locations, or (2) these early protostellar rings are the precursors to the protoplanetary substructure.

Considering path (1), the protostellar ring-like substructures are regions of higher dust densities, and then dust growth will be more efficient (e.g., Gonzalez et al. 2017; Dullemond et al. 2018; Pinilla et al. 2020; Drazkowska et al. 2022; Jiang & Ormel 2023). This would allow early first-generation planets to form at the protostellar ring locations. The planet-disk dynamical interactions can create secondary features, gaps and new rings leading to structured protoplanetary disks (e.g., DSHARP). The location of these new rings would be offset from the original enhancement observed at the protostellar stage as planets would open up gaps and rings separately from where they formed. The early planet formation hypothesis, starting in the protostellar rings, would help alleviate the mass budget problem (e.g., Tychoniec et al. 2020). As observed, protoplanetary disk millimeter dust masses can only reproduce the exoplanet mass budget if the dust-to-planet conversion is 100% efficient or if the disk is continuously replenished with additional material (Manara et al. 2018; Mulders et al. 2021b). So, it was suggested that planet formation starts early, during the protostellar disk phase (Tychoniec et al. 2020).

Alternatively, path (2) is the product of time whereby the underlying mechanism at work forming the protostellar rings continues to impact the disk as it evolves, thus defining the substructures observed at the protoplanetary stage. From our results, the protostellar ring width must decrease over time if these are to become the protoplanetary disk substructure features as shown by the median ring widths. The ring location observed in the protostellar disks could remain fixed if a pressure-bump-induced dust barrier halts radial drift (Carrera et al. 2021). This could allow for the further formation of substructures as the ring presence halts or diminishes radial drift impacting the global disk evolution (e.g., Pinilla et al. 2020). As shown in Figure 9, there is no marked evolution between the protostellar and protoplanetary ring locations as a ratio of disk size. However, if the disk size decreases, as suggested by Hendler et al. (2020), the ring location may also shift. The reasoning explaining why such a ring would narrow between the protostellar and protoplanetary disk stages is not entirely clear yet. Simulations of disk evolution coupled with dust growth and planet-disk interactions are necessary to trace the substructures' evolution. So far, this has been done for evolved protoplanetary disk ring features (e.g., Jiang & Ormel 2021, 2023).

### 4.8. *Limitations*

While we find evidence of substructures toward several protostellar disks, we make several assumptions about the disks. In particular, we assume the emission is axisymmetric and well described by a-priori brightness profiles where the disks are geometrically thin. We do not include disk physics, such as temperature stratification, dust density or optical depth, or grain sizes, which can affect the flux emission at both bands. Additionally, we are fundamentally limited in our description of substructures by the resolution of the observations.

Our axisymmetric disk assumption should mean the Imaginary terms in the *uv* visibilities are zero. Most disks presented here are consistent with zero-value Imag-inary visibilities within the sensitivity uncertainties. IRS 63 and GSS 30 IRS 3 are potential exceptions. In the case of IRS 63, there are low-level Imaginary *uv* profile deviations from zero between 3 and 5$\sigma$ around 500 k$\lambda$.

The models also assume the disks are geometrically thin and in most cases this is a reasonable assumption. The geometrically thin disk approximation is accurate for some evolved protoplanetary disks. For example, Oph 163131 has a vertical thickness of <0.5 au out to 100 au (Villenave et al. 2022). However, at the protostellar stage, edge-on disks have been found to have a larger vertical thickness and possible flaring in millimeter dust emission (Villenave et al. 2020; Lee et al. 2017; Ohashi et al. 2022; Sheehan et al. 2022b; Lin et al. 2023). In our sample, these exceptions are VLA 1623 West (Michel et al. 2022) and GSS 30 IRS 3, where we see residual features that are indicative of a highly inclined, flared disk. Indeed, recent high-resolution observations of GSS 30 IRS 3 show evidence of a flared disk shape (e.g., Ohashi et al. 2023, Santamaria-Miranda et al. in prep.). The observed emission of an inclined flared disk will be impacted by the changes in optical depth and temperature from the disk mid-plane to the outer flared edges (e.g., Ohashi et al. 2022) and there may be non-negligible contributions to the intensity from scattered emission (e.g., Perrin et al. 2015; Yang et al. 2017). While we consider GSS 30 IRS 3 to be a candidate structured disk given our selection criteria, see Appendix A.1, we acknowledge that the disk should be modelled with a flared, radiative transfer model to verify whether such ringed structure truly exists.

Protostellar disks can also be optically thick (see Appendix B.2). The high optical depths signify that emission is absorbed and re-emitted internally within the dust disk such that we do not capture all of the emission equally at both bands. High optical depths and the unknown disk temperature profiles prevent us from calculating accurate protostellar disk dust masses and probing the disk structure at the mid-plane. Longer wavelength observations are sensitive to larger dust populations and are necessary to observe optically thin emission from the disk mid-plane and better characterize internal structures.

The deep sensitivities of the Ophiuchus polarization survey (Sadavoy et al. 2018) from which we obtained the 1.3 mm observations allow us to more accurately probe small emission variations along the disks' radial extents at sub-beam scales (e.g., as done for WL 17 by Gulick et al. 2021). The substructures cannot be seen in the image plane, and the features become apparent purely in the *uv* visibility analysis of these sensitive observations. Between the 1.3 and 0.87 mm data, the more sensitive 1.3 mm observations do a superior job of recovering small emission variations. As we search for the origins of substructures at the protostellar disk stage, high-resolution and high-sensitivity observations are necessary. Nevertheless, this work is the first step in providing key insights into the types of protostellar disks to host substructures and the properties of those substructures.

## 5. SUMMARY AND CONCLUSIONS

We fit ALMA 1.3 and 0.87 mm observations of ten Ophiuchus protostellar disks using simple geometric models to search for evidence of substructure. We use `GALARIO` to model the emission with simple analytic profiles and describe any protostellar disk substructure. The results and interpretations are as follows:

1. Out of ten protostellar sources examined, four have substructure, two appear non-structured, and four are ambiguous disks. Taking the entire protostellar population of Ophiuchus, at least 16% of all the protostellar disks are structured in agreement with the fractions seen in protoplanetary disks. This result suggests that substructures in protostellar disks may be as common as at the protoplanetary stage.
2. The structured protostellar disks tend to be brighter and larger than the non-structured disks. This trend agrees with protoplanetary disk observations, where the massive and largest disks are more likely to host substructures (van der Marel & Mulders 2021).
3. In all the four cases, we identify the substructures as wide Gaussian ring enhancements over a smooth disk profile. We find no evidence of cavities or gaps in any of the disks.
4. The rings are typically wider ($\sigma_R/R_{\rm disk} \sim 0.26$) and have low contrasts ($C < 0.2$) relative to rings observed in protoplanetary disks from the DSHARP survey ($\sigma_R/R_{\rm disk} \sim 0.08$ and $C$ ranges from $0-1$). This difference points to two potential pathways: (1) the protostellar rings are the sites where planets will form, and the ring-gap pairs seen in protoplanetary disks are a secondary feature, or (2) the protostellar rings will evolve over the disk lifetime to become narrow and with higher contrast.
5. The presence of shallow, broad (relative to disk size) ring enhancements in the protostellar disks without gaps does not match most of the current substructure formation theory or simulations in protoplanetary disks. Therefore, it is necessary to develop additional, new, varied models and theories applied to substructure formation and evolution in protostellar disks.

6. The disk substructures in Ophiuchus are also morphologically distinct from those seen in protostellar disks in Orion, which could be due to the type of protostar properties or the mechanism that caused the observed substructures.

The presence of substructures in multiple disks at the protostellar disk phase provides new insights into the origins of disk substructure. The dust rings are ideal for early dust growth and planet formation. This provides a solution to the early start for planet formation and guides the first-generation planet formation description. Supposing planets form in these original protostellar rings, these will clear a dust lane at that location and induce new substructures at different radii, which we likely observe at the protoplanetary disk stage.

The resolution of the observations limits the precise characterization of the substructure features. However, the high sensitivity of these data still allows us to find and generally describe the variety of substructures from the *uv* visibilities. Future high-resolution observations and more detailed radiative transfer modelling are required to improve the description of protostellar disk substructures and increase the sample size. Combined, they can provide valuable results regarding the substructure properties and connect these with dust growth, planet formation, and disk evolution models.

*Software:* `CASA` (McMullin et al. 2007), `GALARIO` (Tazzari et al. 2018), `emcee` (Foreman-Mackey et al. 2013), `matplotlib` (Hunter 2007), `corner` (Foreman-Mackey 2016), `APLpy` (Robitaille & Bressert 2012), `frank` (Jennings et al. 2020).

*Acknowledgments:* We thank an anonymous referee for their insightful comments and suggestions. AM and SIS acknowledge support from the Natural Science and Engineering Research Council of Canada (NSERC), RGPIN-2020-03981. LWL acknowledges support from NSF AST-1910364 and NSF AST-2108794. ALMA is a partnership of ESO (representing its member states), NSF (USA) and NINS (Japan), together with NRC (Canada) and NSC and ASIAA (Taiwan), and KASI (Republic of Korea), in cooperation with the Republic of Chile. The Joint ALMA Observatory is operated by ESO, AUI/ NRAO, and NAOJ. This paper uses the following ALMA data: 2015.1.01112.S, 2015.1.00741.S, and 2015.1.00084.S.

# APPENDIX

## A. STRUCTURE DETAILS

### A.1. *Residuals after a smooth disk subtraction from structured sources*

In Figure 11 we show the imaged residuals of the structured protostellar disks after a smooth disk (PLCT) model has been subtracted. The symmetric residuals along the major axis are indicative of the ring feature for these inclined disks. While we do not model spirals, from the smooth-subtracted imaged residuals, we do not find any evidence of spiral features.

GSS 30 IRS 3 is possibly also a flared disk (Ohashi et al. 2023), and given the disk's high inclination, the possible ring signature may be conflated with emission from the flaring. Nevertheless, we categorize this disk as structured given that the statistical assessment from the ΔAIC and ΔBIC provide evidence in favor of the source being structured rather than smooth or flared. From Table 7, the PLCT, ring disk model is strongly and positively favored in contrast to a smooth PLCT disk model at 1.3 mm for the ΔAIC and ΔBIC, respectively. We cannot make this comparison at 0.87 mm due to the lower sensitivities at the source.

As highly inclined protostellar disks are observed at higher resolution and sensitivities, more flared structures are being observed (e.g., Lee et al. 2017; Ohashi et al. 2022; Sheehan et al. 2022b; Lin et al. 2023) and thus our sample of ambiguous disks, for which the source of excess emission is uncertain, could be good candidates for follow up observations to search for and quantify millimeter-dust disk flaring and its evolution.

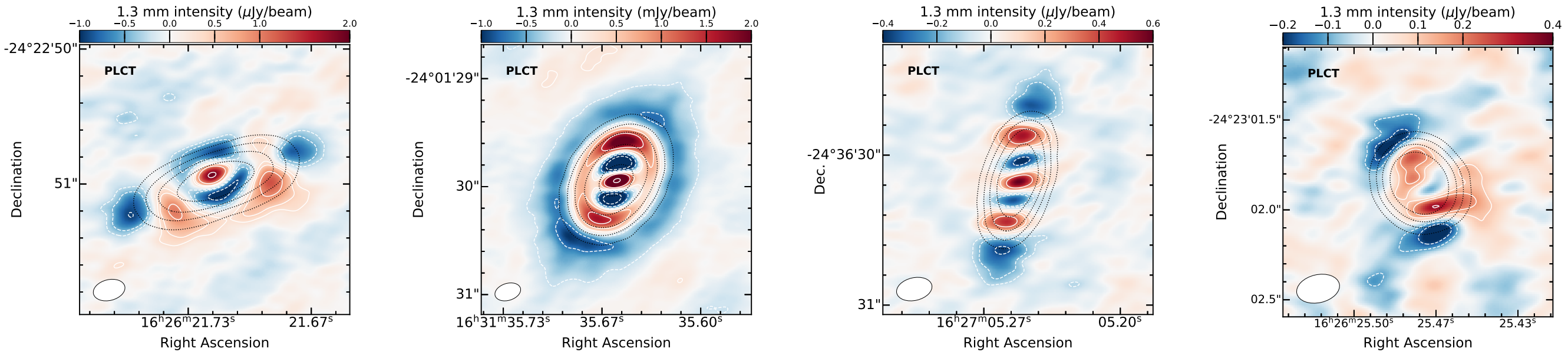

**Figure 11.** Imaged residuals of the four structured protostellar disks after subtracting a smooth disk (PLCT) model, from left to left to right GSS 30 IRS 3, IRS 63, Oph-emb-6, Oph-emb-9.

### A.2. *Offset Gaussian in IRS 44*

We present the offset Gaussian component needed for the mdoels of IRS 44. When fitting a PLCT and Gaussian envelope model, we find excess residual emission along the Eastern edge of the disk, ∼40 au away from the disk center see Figure 12, while $R_{90\%}$∼35 au. The residual is consistent at both wavelengths, 1.3 and 0.87 mmm, implying this is a real feature. We find that a Gaussian function can describe this excess emission well such that the complete model fits the data well, leaving $\leqslant 6\sigma_{1.3\mathrm{mm}}$ and $<3\sigma_{0.87\mathrm{mm}}$ residuals, see Figures 3 and 15. Although the offset Gaussian adds four new free parameters, both the AIC and BIC strongly favor the inclusion of this component at 1.3 mm. At 0.87 mm, the AIC favors the inclusion of the offset Gaussian, but the BIC, which more heavily penalises additional parameters, favors the simpler PLCT and Gaussian envelope model. Artur de la Villarmois et al. (2022) observed this source in sulfur emission (six lines of $SO_2$) at 0.1″. From these observations, they find shocked accretion taking place south of the disk and infalling-rotating motions within the disk region <30 au and colder less energetic gas at ∼400 au but no gas tracers highlight any anomaly to the east at ∼40 au. This is thus an interesting source for follow-up observations to elucidate the reason for this excess continuum emission east of the disk.

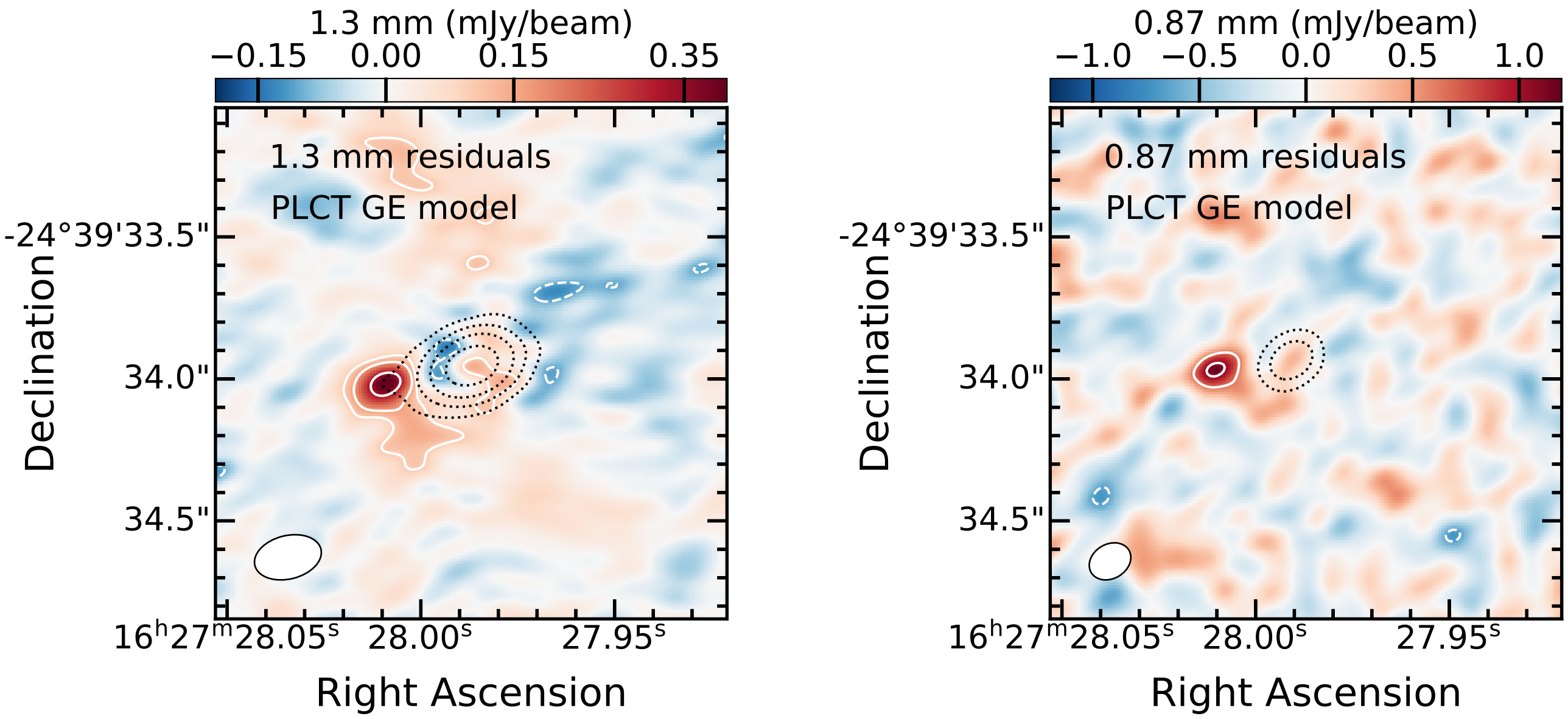

**Figure 12.** IRS 44 imaged residuals after subtracting a best-fit PLCT and Gaussian Envelope model from the observations. The white contours $\pm 3, 5, 10\sigma$ residuals and black dotted contours are for the observed soure emission with $20, 50, 100, 200\sigma$.

## B. 0.87 MM RESULTS

### B.1. *0.87 mm Figures*

We present the 0.87 mm results in Figures 13, 15, and 14. The 0.87 mm data are analyzed and presented like the higher sensitivity 1.3 mm data in Section 3. We provide the resulting best-fit parameters for both wavelengths in Table 5. The notable differences between the favored best-fit disk models between the 0.87 and 1.3 mm data concerns GSS 30 IRS 3 and IRS 63. The remaining seven targets fit with the same disk models at both wavelengths. We further note that the 0.87 mm data are less sensitive than the 1.3 mm data (except for VLA 1623-West), and therefore the residuals tend to be less significant at 0.87 mm.

At 0.87 mm, GSS 30 IRS 3 is far off the primary beam center. As a result, the noise is much higher such that the disk has a peak SNR of 26 at 0.87 mm compared to 429 at 1.3 mm (see Figure 13). Therefore, the ring component cannot be identified or fit at 0.87 mm. The 0.87 mm data of IRS 63 are not sensitive enough at the larger $uv$-distances to model the inner Gaussian feature detected at 1.3 mm. Therefore, the 0.87 mm model excludes this component.

### B.2. *Spectral index*

To study the dust properties of the protostellar disks, we measure the spectral index, $\alpha_{mm}$, with the 1.3 and 0.87 mm data assuming $S \sim \nu^{\alpha}$. We clean the observations with uniform weighting across the same $uv$ range for each source and smoothed each map by the beam of the other so that the two images have a common resolution and beam shape. We center the observations according to the peak image plane emission from a 2D Gaussian fit. We do not use the `GALARIO` model centers as those occasionally vary from the emission center since we model more than the disk emission in multiple cases, e.g., including large-scale Gaussian envelopes.

Using the `immath` task in `CASA`, we evaluate the pixel-by-pixel spectral index map from the two wavelength observations. We mask emission $< 10\sigma$ at both wavelengths to focus on the main disk emission. For GSS 30 IRS 3, the 0.87 mm data are not sensitive enough to provide a reliable detection above $10\sigma_{0.87\mathrm{mm}}$, so no spectral index is evaluated for this source.

Figure 16 shows the spectral index maps for the protostellar disks. A value of $\alpha = 2$ is expected if the disk is optically thick (e.g., the dust emission follows a black body function instead of a modified black body function) or has very large dust grains. The steeper index at larger radii would indicate less optically thick emission.

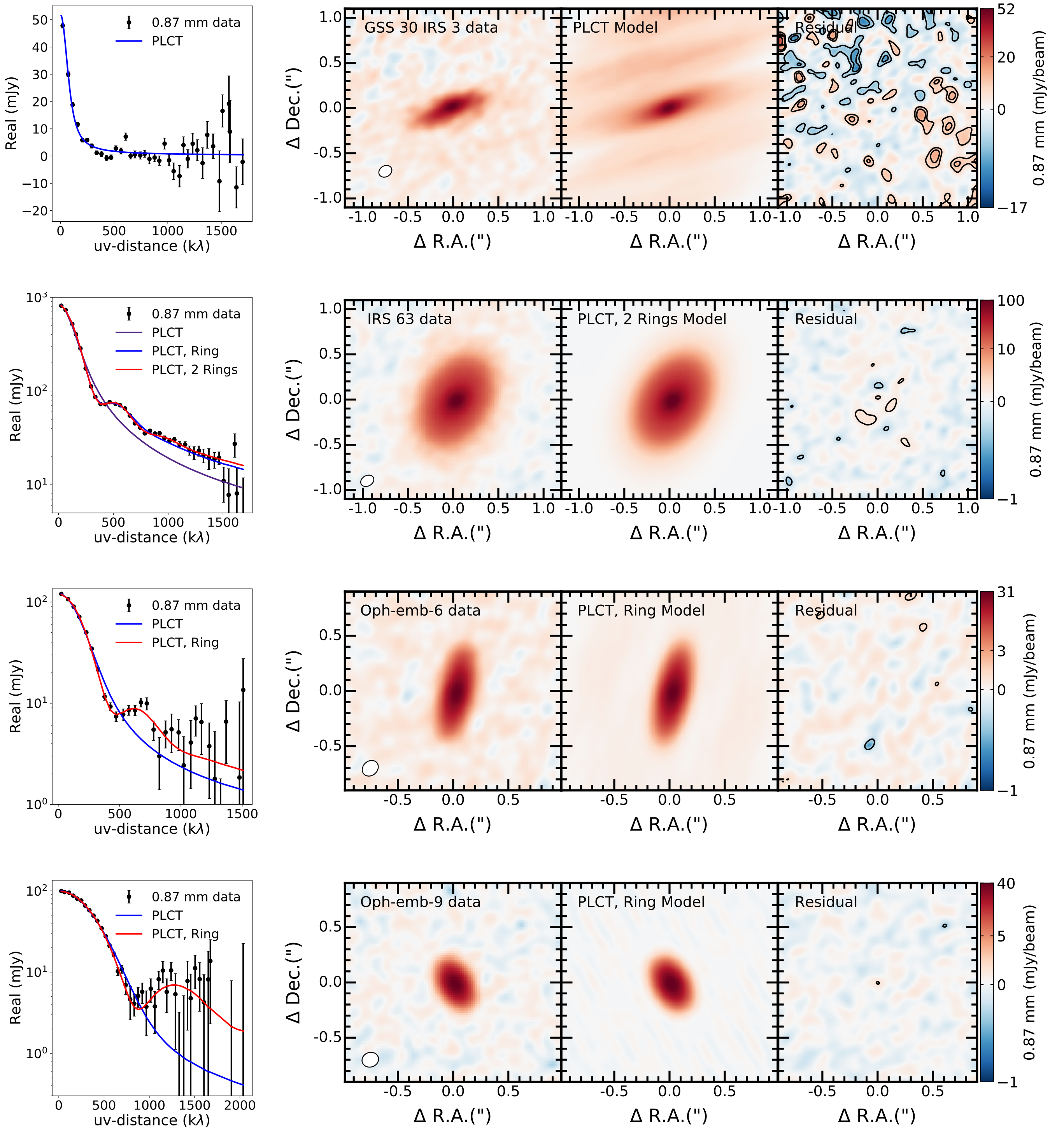

**Figure 13.** Same as Figure 1 for 0.87 mm data of the structured disks. We show the $uv$ data in 45 k$\lambda$ bins.

Following Michel et al. (2022), we measured the spectral indices using the 1.3 and 0.87 mm data, assuming 10% flux calibration uncertainties for each wavelength. We find typical values of $\alpha_{mm} \sim 2$ toward the center and $\alpha_{mm}$ steepening at larger radial extents, implying that the dust emission may be optically thick at the centre of each disk and transitioning to optically thin at the disk edge. The errors on $\alpha_{mm}$ from the 1.3 and 0.87 mm data alone are typically $\pm 0.5$, implying that some disks (e.g., Elias 29 and IRS 37-A) could be consistent with a flat spectral index at all radii within uncertainties. IRS 44 and Oph-emb-1 show a gradient in spectral along the disk minor axis. However, this gradient may be due to a misalignment of the disk centers based on the 2D Gaussian fits. Both sources have substantial envelope components that could affect our measurement of a disk's central position.

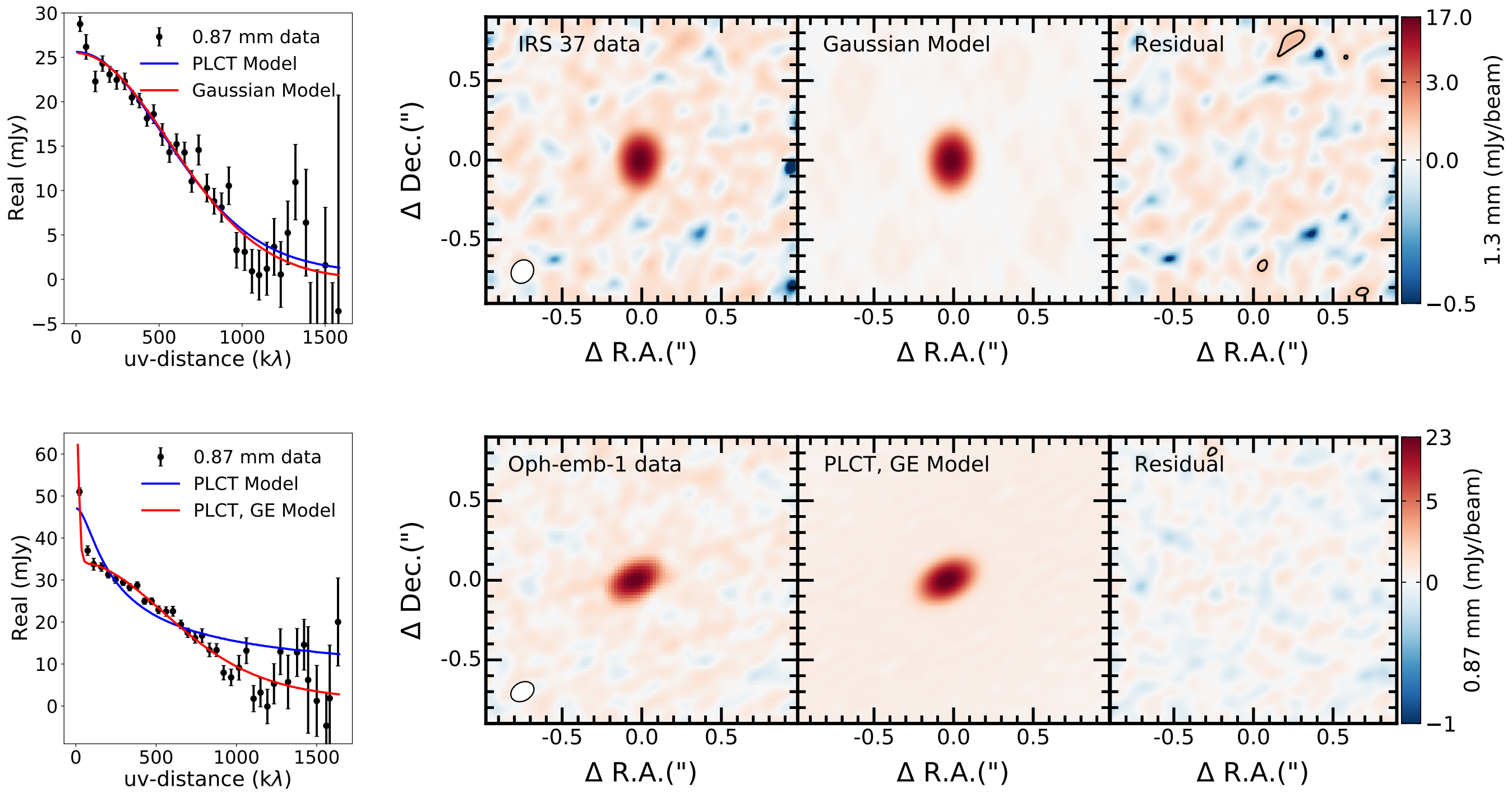

**Figure 14.** Same as Figure 2 for 0.87 mm data of the non-structured disks. We show the $uv$ data in 45 k$\lambda$ bins.

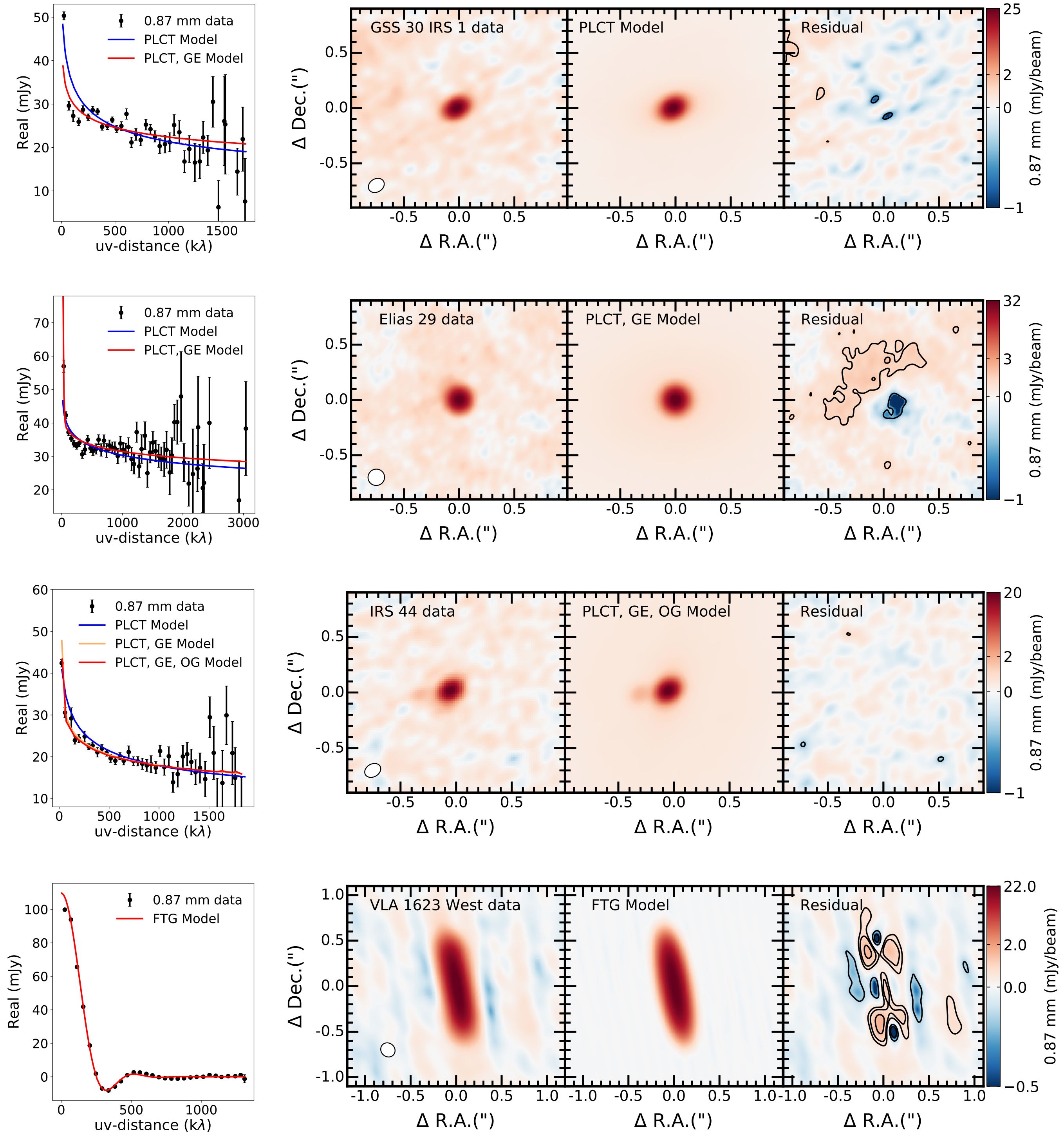

**Figure 15.** Same as Figure 3 for 0.87 mm data of the ambiguous disks. We show the $uv$ data in 45 k$\lambda$ bins.

## C. 1D BEST-FIT MODEL BRIGHTNESS PROFILES

In Figures 17, 18, and 19, we present the disk and envelope model components and the cumulative best-fit model as brightness profiles as a function of radius.

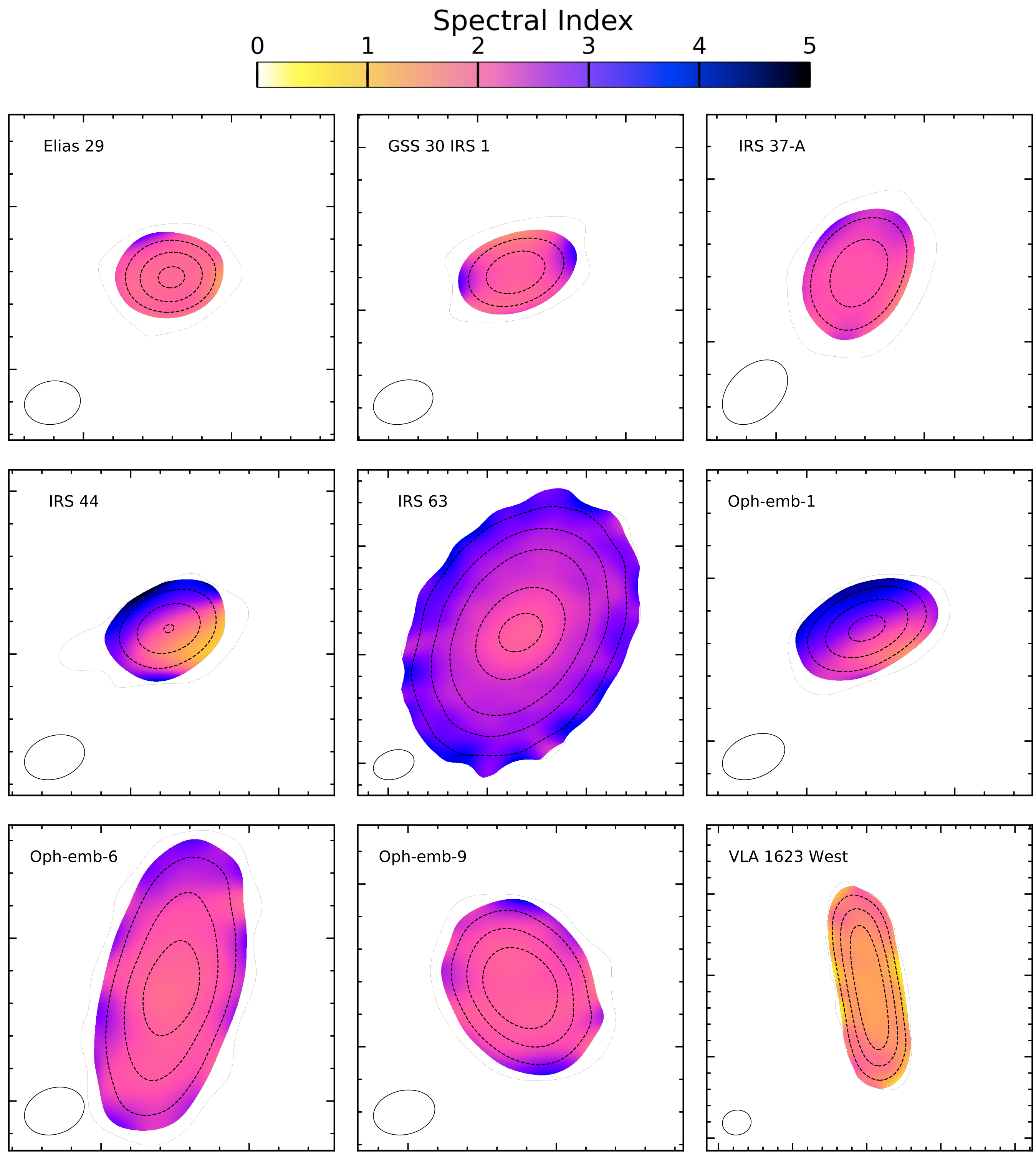

**Figure 16.** Spectral index maps of the protostellar disk sample. GSS 30 IRS 3 does not have a spectral index map since the 0.87 mm observations are so far off the primary beam center, these are not adequate to reliably evaluate a spectral index.

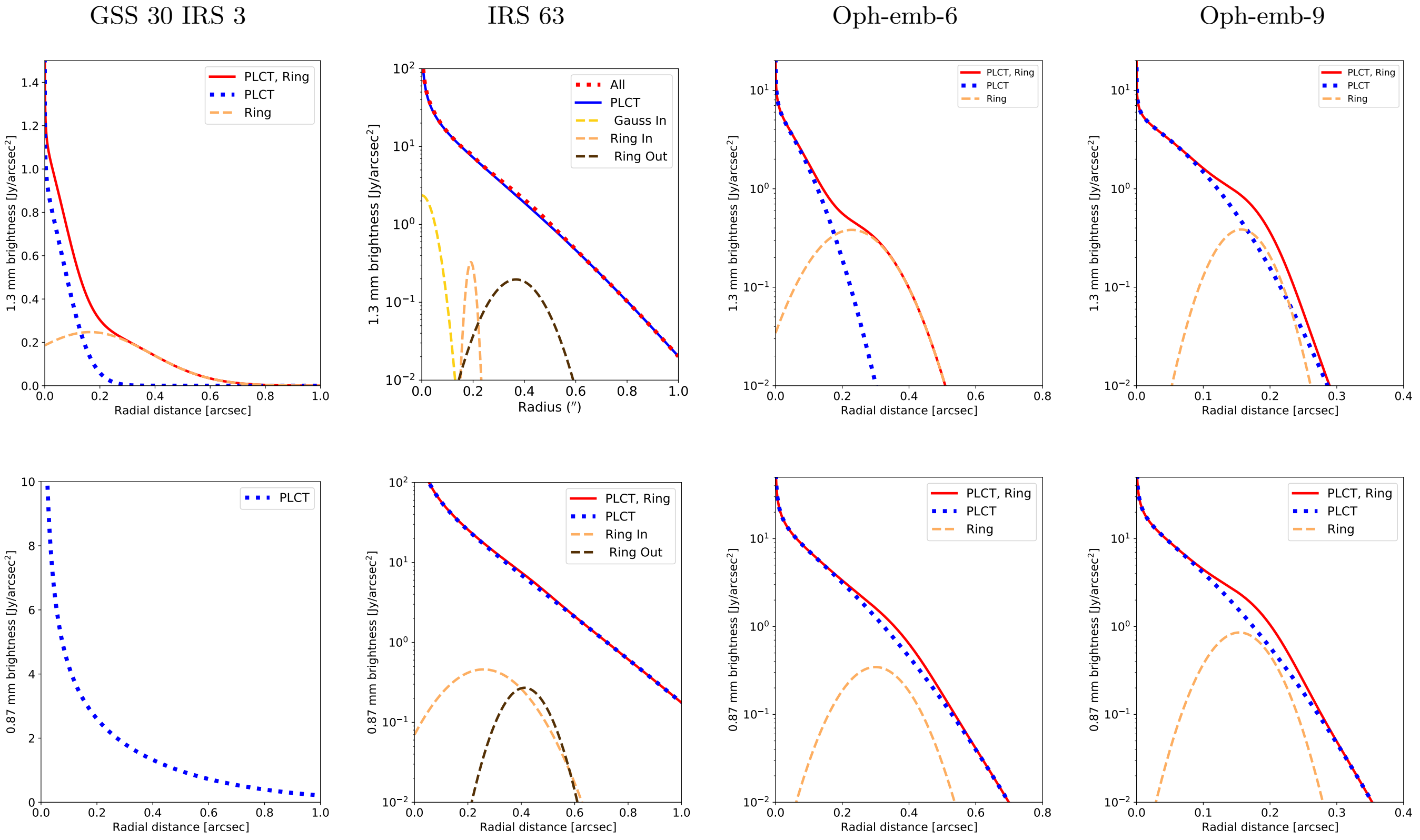

**Figure 17.** Structured disks best-fit model intensity profiles and their components. From left to right, GSS 30 IRS 3, IRS 63, Oph-emb-6, and Oph-emb-9. The top row shows the 1.3 mm best-fit models, and the bottom row shows the 0.87 mm models.

## D. STATISTICAL TEST RESULTS

Tables 7, 8, and 9 show the results from the statistical tests for the model variations applied to the 1.3 and 0.87 mm data. For the statistical tests performed on the various models fit to VLA 1623 West, we refer the reader to Table 2 in Michel et al. (2022). $\Delta$AIC and $\Delta$BIC provide a quantitative comparison and statistically favored model and are evaluated as $\Delta\mathrm{AIC} = \mathrm{AIC}_0 - \mathrm{AIC}_1$, where $\mathrm{AIC}_0$ is the reference model and $\mathrm{AIC}_1$ the model being evaluated against the reference. The same method is applied for $\Delta$BIC. The reference models used are labelled in the tables.
If $\Delta$AIC and $\Delta$BIC are positive within the following ranges, then:

- 3 to 20, there is positive evidence in favour of $\mathrm{AIC}_1$ or $\mathrm{BIC}_1$,
- 20 to 150, there is strong evidence in favour of $\mathrm{AIC}_1$ or $\mathrm{BIC}_1$,
- $> 150$, there is decisive evidence in favour of $\mathrm{AIC}_1$ or $\mathrm{BIC}_1$.

If $\Delta$AIC and $\Delta$BIC are negative within the following ranges, then:

- -3 to -20, there is positive evidence in favour of $\mathrm{AIC}_0$ or $\mathrm{BIC}_0$, the reference model,
- -20 to -150, there is strong evidence in favour of $\mathrm{AIC}_0$ or $\mathrm{BIC}_0$, the reference model,
- $< -150$, there is decisive evidence in favour of $\mathrm{AIC}_0$ or $\mathrm{BIC}_0$, the reference model.

In cases where $\Delta$AIC and $\Delta$BIC range from -3 to 3, neither model is statistically favored.

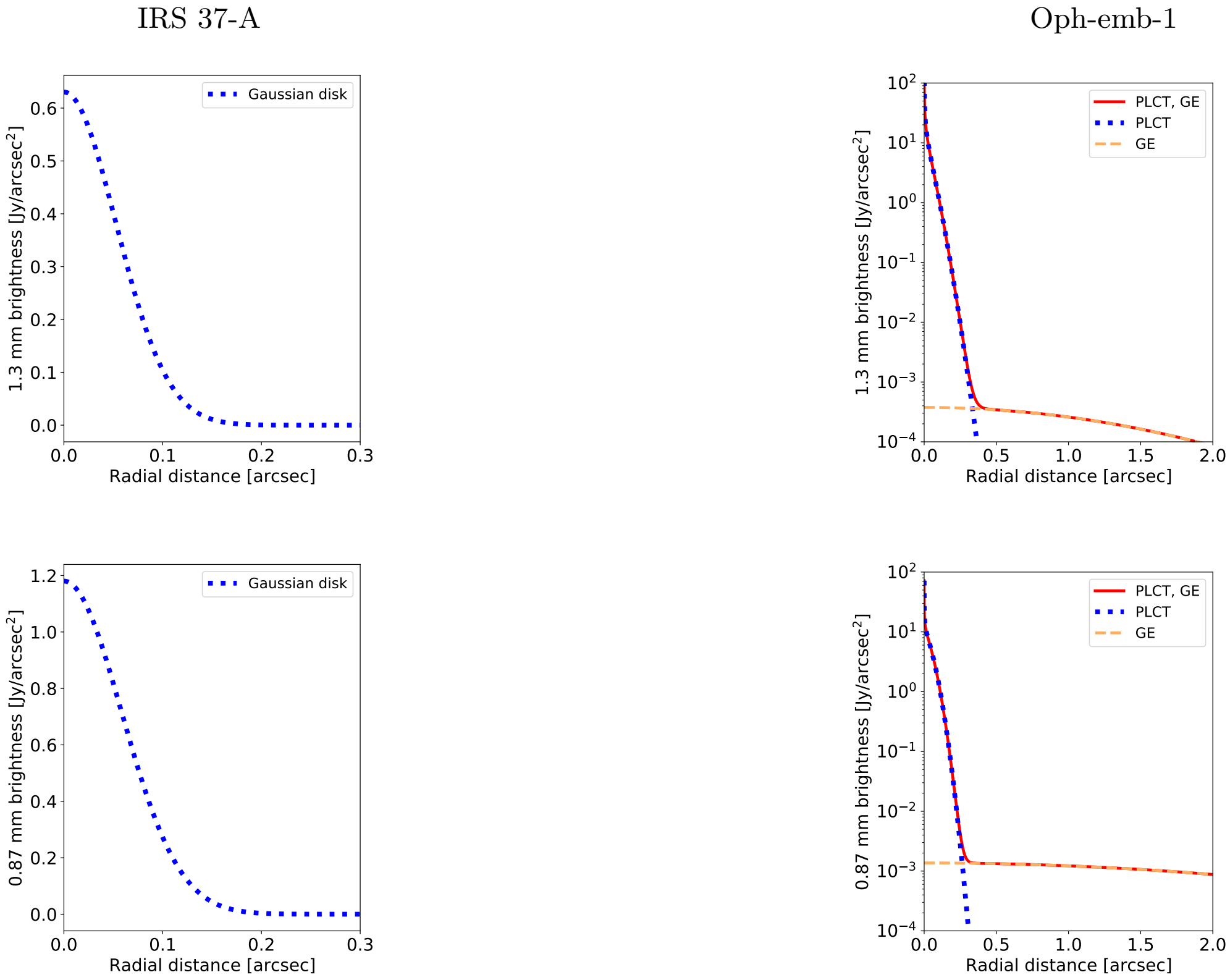

**Figure 18.** Same as Figure 17 for the non-structured disks, from left to right IRS 37-A and Oph-emb-1.

**Table 7.** Model statistical comparison for structured protostellar disks.

| Source | GSS 30 IRS 3 | | IRS 63 | | Oph-emb-6 | | Oph-emb-9 | |
|---|---|---|---|---|---|---|---|---|
| $\lambda$ | 1.3 mm | 0.87 mm | 1.3 mm | 0.87 mm | 1.3 mm | 0.87 mm | 1.3 mm | 0.87 mm |
| | | | | $\Delta$AIC | | | | |
| FTG | -880 | -25 | -28394 | 7674 | -912399 | -225 | -164 | -35 |
| PLCT | -43 | 8599[a] | -4304 | -2096 | -434 | -51 | -140 | -20 |
| PLCT, GR | 55725[a] | ... | -33 | -8 | 49342[a] | 2815[a] | 28980[a] | 2791[a] |
| PLCT, 2GR | ... | ... | -162 | 4842[a] | ... | ... | ... | ... |
| PLCT, 2GR, IG | ... | ... | 139898[a] | ... | ... | ... | ... | ... |
| | | | | $\Delta$BIC | | | | |
| FTG | -853 | -24 | -28322 | -7635 | -912372 | -205 | -137 | -9 |
| PLCT | -16 | 8645[a] | -4232 | -2057 | -407 | -31 | -113 | -1 |
| PLCT, GR | 55815[a] | ... | 12 | 12 | 49432[a] | 2880[a] | 29070[a] | 2861[a] |
| PLCT, 2GR | ... | ... | -144 | 4927[a] | ... | ... | ... | ... |
| PLCT, 2GR, IG | ... | ... | 140033[a] | ... | ... | ... | ... | ... |

*Notes:*

a Reference model ($AIC_0$, $BIC_0$) to which other models are compared, i.e., $\Delta AIC = AIC_0 - AIC_1$.

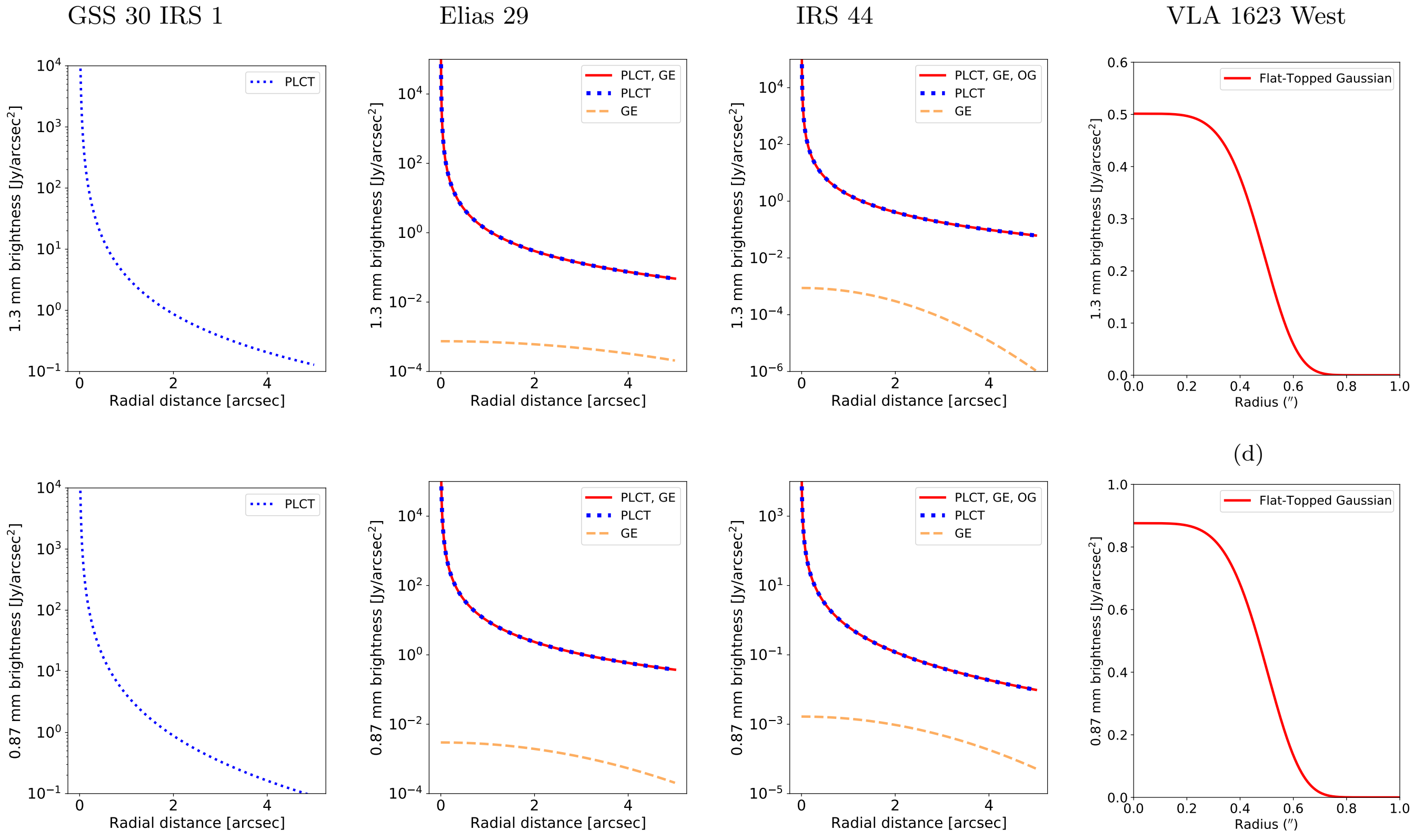

**Figure 19.** Same as Figure 17 for the ambiguous disks, from left to right GSS 30 IRS 1, Elias 29, IRS 44, and VLA 1623 West. For IRS 44, we do not show the individual offset Gaussian component on this brightness versus radius plot as it is an isolated, localized feature which does not exist all around the disk at a particular radius, it is localized.

**Table 8.** Model statistical comparison for non-structured protostellar disks.

| Source | IRS 37-A | | Oph-emb-1 | |
|---|---|---|---|---|
| $\lambda$ | 1.3 mm | 0.87 mm | 1.3 mm | 0.87 mm |
| | AIC | | | |
| PLCT | -3 | -2 | -102 | -157 |
| PLCT, GE | 5 | 5 | 36036[a] | 2720[a] |
| Gaussian | 34421[a] | 2693[a] | ... | ... |
| | BIC | | | |
| PLCT | -12 | -9 | -84 | -143 |
| PLCT, GE | -21 | -15 | 36117[a] | 2779[a] |
| Gaussian | 34475[a] | 2732[a] | ... | ... |

*Notes:*

a Reference model ($AIC_0$, $BIC_0$) to which other models are compared, i.e., $\Delta AIC = AIC_0 - AIC_1$.

**Table 9.** Model statistical comparison for ambiguous protostellar disks. For VLA 1623 West refer to Table 2 in Michel et al. (2022).

| Source | GSS 30 IRS 1 | | Elias 29 | | IRS 44 | |
|---|---|---|---|---|---|---|
| $\lambda$ | 1.3 mm | 0.87 mm | 1.3 mm | 0.87 mm | 1.3 mm | 0.87 mm |
| | | | AIC | | | |
| PLCT | -6 | 45 | -896 | -44 | -347 | -60 |
| PLCT, GE | 54205[a] | 5463[a] | 39043[a] | 2450[a] | -138 | -12 |
| PLCT, OG | ··· | ··· | ··· | ··· | 38110[a] | 2888[a] |
| | | | BIC | | | |
| PLCT | 12 | 59 | -878 | -32 | -293 | -20 |
| PLCT, GE | 54286[a] | 5522[a] | 39124[a] | 2394[a] | -102 | 14 |
| PLCT, OG | ··· | ··· | ··· | ··· | 38227[a] | 2973[a] |

*Notes:*

a Reference model ($AIC_0$, $BIC_0$) to which other models are compared, i.e., $\Delta AIC = AIC_0 - AIC_1$.